\documentclass[12pt]{iopart}
\usepackage{graphicx}
\usepackage{dcolumn}
\usepackage{bm}
\usepackage{color}
\usepackage{amsthm, amssymb, amsfonts}
\usepackage{ulem}
\usepackage{setspace}
\newcommand{\da}{^\dagger}

\begin{document}

\title[]{Theoretical study of reflection spectroscopy for superconducting quantum parametrons}

\author{S. Masuda$^{1}$, A. Yamaguchi$^{2,3}$, T. Yamaji$^{2,3}$, T. Yamamoto$^{2,3}$, T. Ishikawa$^{1,3}$, Y. Matsuzaki$^{1,3}$ and S. Kawabata$^{1,3}$}
\address{$^{1}$ Research Center for Emerging Computing Technologies (RCECT), National Institute of Advanced Industrial Science and Technology (AIST), 1-1-1, Umezono, Tsukuba, Ibaraki 305-8568, Japan
}
\address{$^{2}$ System Platform Research Laboratories, NEC Corporation, Kawasaki, Kanagawa 211-0011, Japan}
\address{$^{3}$ NEC-AIST Quantum Technology Cooperative Research Laboratory, National Institute of Advanced Industrial Science and Technology (AIST), Tsukuba, Ibaraki 305-8568, Japan}

\vspace{10pt}

\begin{abstract}
Superconducting parametrons in the single-photon Kerr regime, also called KPOs,  have been attracting increasing attention in terms of their applications to quantum annealing and universal quantum computation.
It is of practical importance to obtain information of superconducting parametrons operating under an oscillating pump field.
Spectroscopy can provide information of a superconducting parametron under examination, such as energy level structure, and also useful information for calibration of the pump field. 
We theoretically study the reflection spectroscopy of superconducting parametrons, and develop a method to obtain the reflection coefficient. 
We present formulae of the reflection coefficient, the nominal external and the internal decay rates, and examine the obtained spectra.
It is shown that the difference of the populations of energy levels manifests itself as a dip or peak in the amplitude of the reflection coefficient, and one can directly extract the coupling strength between the energy levels by measuring the nominal decay rates when the pump field is sufficiently large.
\end{abstract}

%
%
%
%
%

\section{Introduction}
Classical parametric phase-locked oscillators \cite{Onyshkevych1959}, called parametrons \cite{Goto1959},
were operated as classical bits in digital computers in the 1950s and 1960s.
Recently, parametrons in the single-photon Kerr regime \cite{Kirchmair2013}, where the nonlinearity is larger than the decay rate, have been attracting much attention in terms of their applications to quantum information processing. 
Parametrons were applied to the qubit readout~\cite{Yamamoto2014,Yamamoto2016} in the circuit QED architecture, a promising platform of quantum information processing \cite{You2005,You2007,You2011,Gambetta2017,Wendin2017,Krantz2019,Gu2019,Blais2020}.
Quantum annealing \cite{Goto2016,Nigg2017,Puri2017,Kanao2021}
and universal quantum computation \cite{Goto2016b}, which utilize the quantum nature of parametrons in a superconducting circuit, have been proposed.
More recently, bias-preserving gates~\cite{Puri2020} and single-qubit operations~\cite{Grimm2020} were studied theoretically and experimentally, and the exponential increase of the bit-flip time with the cat size was observed \cite{Lescanne2020}.

A parametron in the single-photon Kerr regime is operated by an oscillating pump field.
Therefore, to obtain information of the parametron under the pump field and to calibrate the amplitude of the pump field are practically important. 
In previous studies, state tomography of parametrons using the power spectrum density were demonstrated~\cite{Wang2019}; energy differences between either of the two highest energy levels and a lower energy level of a parametron were measured by mapping the parametron to a Fock qubit~\cite{Grimm2020}; and microwave responses of parametrons without a pump field were experimentally investigated in a wide range of the Kerr nonlinearity \cite{Yamaji2021}.
However, theories of reflection and transmission spectroscopy of parametrons have not been developed in spite of the fact that they are important and routinely applied to resonators to examine the energy level structure and their quality. 
Spectroscopy will provide useful information of superconducting parametrons under examination, such as energy level structure.

Recently, a fast and accurate gate operation of a parametron utilizing energy levels outside of the qubit space was proposed~\cite{Kanao2021b}. In the method, the couplings between either of the two highest levels and other lower levels induced by a drive pulse are essential. For implementation of the technique, it is important to measure the energy differences and the couplings between relevant levels.

In this paper, we develop a method to obtain the reflection coefficient of a pumped superconducting parametron.
We present simple formulae of the reflection coefficient and the nominal decay rates
\footnote{
The external and the internal decay rates or the quality factors of resonators are routinely measured by spectroscopic methods. They are obtained by fitting the spectra to an analytic form. The nominal decay rates of a parametron can be obtained in the same manner as a resonator, and can provide information of the measured parametron as explained in Sec~\ref{Weak input field limit}.}.
It is shown that one can directly extract the amplitude of the coupling coefficients between energy levels by measuring the nominal decay rates when the pump field is sufficiently large.
Moreover, we show that the nominal internal decay rate increases with the pump strength and eventually exceeds the nominal external decay rate even if the original internal decay rate of the parametron without a pump field is negligible.
Our method of spectroscopy does not require pulsed operations of a parametron, and can be implemented with a standard reflection-measurement setup routinely used for superconducting circuit QED architectures.

\section{Model}
\begin{figure}[h!]
\begin{center}
\includegraphics[width=10cm]{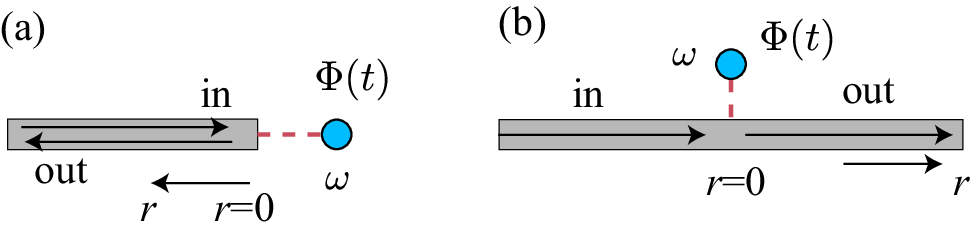}
\end{center}
\caption{
Schematic of the system. 
(a) A parametron (blue circle) is attached to a transmission line, where an incoming and an outgoing microwaves propagate.
An external oscillating magnetic flux $\Phi(t)$ is used to pump the parametron. 
$\omega$ is the resonance frequency of the parametron when no pump filed is applied.
The parametron is located at $r=0$.
(b) Effective model of (a). In the theory in section~\ref{Reflection coefficient 3 29 21}, the negative and positive regions are assigned to the incoming and outgoing fields, respectively. 
}
\label{model_2_27_18}
\end{figure}

We develop a theory to obtain the reflection coefficient for parametrons.
Our method is similar to that in Ref.~\cite{Koshino2013} developed for a driven three-level system.
We consider a system composed of a parametron attached to a transmission line (TL) as depicted in figure~\ref{model_2_27_18}(a).
An incoming and an outgoing microwaves propagate in the same TL.
The parametron is pumped by an external oscillating magnetic flux $\Phi(t)$.

Hamiltonian of the system is given by
\begin{eqnarray}
\frac{\mathcal{H}}{\hbar} &=&   \omega a^\dagger a - \frac{\chi}{12}(a+a^{\dagger})^4 + 2\beta(a + a^{\dagger})^2 \cos\omega_p t\nonumber \\ 
&& +   \int_{0}^{\infty} dk \Big{[} v_b k b_{k}^\dagger b_{k} +
 \sqrt{\frac{v_b\kappa_{\rm ex}}{2\pi}}(a^\dagger b_{k} +  b_{k}^\dagger a) \Big{]}\nonumber \\ 
&& +   \int_{0}^{\infty} dk \Big{[} v_c k c_{k}^\dagger c_{k} +
 \sqrt{\frac{v_c\kappa_{\rm int}}{2\pi}}(a^\dagger c_{k} +  c_{k}^\dagger a) \Big{]},
\label{H_6_6_17}
\end{eqnarray}
where the first line is Hamiltonian of the parametron under a pump field~\cite{Wang2019} (see \ref{Hamiltonian of a parametron} for derivation), and the second line is Hamiltonian of the eigenmodes of the TL and the coupling between the modes and the parametron. 
The third line is Hamiltonian of the eigenmodes representing a loss channel and their coupling to the parametron. The decay to the channel represents the internal decay of the parametron.
The second term in the right-hand side of equation~(\ref{H_6_6_17}) gives rise to an anharmonic term in a rotating frame as shown later.
The third term originates from the pump field $\Phi(t)$.
Here, $\beta$ and $\omega_p$ are the amplitude and the angular frequency of the pump field, respectively.
The annihilation operator of the parametron and those of the eigenmodes of the TL (loss channel) with the wave number $k$ are denoted by $a$ and $b_k$ ($c_k$), respectively. 
The decay rate to the TL (loss channel) and the phase velocity in the TL (loss channel) are denoted by $\kappa_{\rm ex}$ ($\kappa_{\rm int}$) and $v_b$ ($v_c$), respectively.
Hereafter, we assume $v_c=v_b$.

\section{Reflection coefficient}
\label{Reflection coefficient 3 29 21}
In this section, we show the method to calculate the reflection coefficient of a parametron.
In our method, we use the input-output relation. 
We derive the input-output relation to make this paper self-contained, although the derivation is based on a standard approach and can be found, e.g., in~Ref. \cite{Yamamoto2016}. 
For a parametron, energy eigenstates in a rotating frame are important because two of them are utilized as qubit states. We develop a method to obtain the reflection coefficient for a parametron using its energy eigenstates as a basis in section~\ref{Representation of reflection coefficient with density matrix elements}. 
This method enables one to obtain the reflection coefficient without integrating master equation.

The Heisenberg equation of motion of $b_{k}$ is represented as
\begin{eqnarray}
\frac{d}{dt}b_{k}(t) = -i v_b k b_{k}(t) - i\sqrt{\frac{v_b \kappa_{\rm ex}}{2\pi}} a(t).
\label{Hei_9_23_20}
\end{eqnarray}
A formal solution of equation~(\ref{Hei_9_23_20}) is 
\begin{eqnarray}
b_{k}(t) = b_{k}(0)e^{-ikv_bt} - i\sqrt{\frac{v_b \kappa_{\rm ex}}{2\pi}}\int_0^t d\tau a(\tau) e^{ikv_b(\tau-t)},
\label{blk_6_28_17}
\end{eqnarray}
where $t>0$.
We formally extend the lower limit of $k$ to $-\infty$ in order to introduce the real-space represention
of the field operator defined by
\begin{eqnarray}
\tilde{b}_{r}(t) = \frac{1}{\sqrt{2\pi}} \int_{-\infty}^{\infty}dk e^{ikr}b_{k}(t),
\label{br_6_28_17}
\end{eqnarray}
where $r$ runs over $-\infty < r < \infty$.
The negative and positive regions are assigned to the incoming and outgoing fields, respectively, as depicted in 
figure~\ref{model_2_27_18}(b).
Thus, the input field operator $\tilde{b}_{r}^{(\rm in)}$ and the output field operator $\tilde{b}_{r}^{(\rm out)}$ are represented as 
\begin{eqnarray}
\tilde{b}_{r}^{(\rm in)}(t) = \tilde{b}_{-r}(t),\nonumber\\
\tilde{b}_{r}^{(\rm out)}(t) = \tilde{b}_{r}(t).
\label{b_in_out_6_28_17}
\end{eqnarray}
The introduction of the real-space representation has been validated in reference~\cite{Banacloche2013}.
Using equations~(\ref{blk_6_28_17}) and (\ref{br_6_28_17}), we can obtain 
\begin{eqnarray}
\tilde{b}_{r}(t) = \tilde{b}_{r-v_bt}(0) - i \sqrt{\frac{\kappa_{\rm ex}}{v_b}} \theta(r)\theta(v_bt-r) a(t-r/v_b),
\label{br_2_18_21}
\end{eqnarray}
where $ \theta$ is the Heaviside step function.
Using equation~(\ref{b_in_out_6_28_17}) with $r=0$ in equation~(\ref{br_2_18_21}), we obtain
the input-output relation,
\begin{eqnarray}
\tilde{b}_{0}^{(\rm out)}(t) = \tilde{b}_{v_bt}^{(\rm in)}(0) - \frac{i}{2}\sqrt{\frac{\kappa_{\rm ex}}{v_b}} a(t),
\label{in_out_12_4_18}
\end{eqnarray}
where we used $\theta(0)=1/2$.
The Heisenberg equation of motion of $a$ with equations~(\ref{br_6_28_17}), (\ref{b_in_out_6_28_17}) and (\ref{in_out_12_4_18}) leads to 
\begin{eqnarray}
\frac{d}{dt}a &=& \Big{(} -i\omega - \frac{\kappa_{\rm tot}}{2} \Big{)} a
+ i \frac{\chi}{3}(a + a^\dagger)^3 
- i 4 \beta \cos(\omega_p t) (a + a^\dagger)\nonumber\\
&&- i \sqrt{v_b\kappa_{\rm ex}} {\tilde{b}}_{v_bt}^{(\rm in)}(0) - i \sqrt{v_b\kappa_{\rm int}} {\tilde{c}}_{v_bt}^{(\rm in)}(0),
\label{a_6_6_17_2}
\end{eqnarray}
where $a$ abbreviates $a(t)$, and we formally extended the lower limit of $k$ to $-\infty$ in equation~(\ref{H_6_6_17}). 
Here, $\tilde{c}_{v_bt}^{(\rm in)}$ is the counterpart of $\tilde{b}_{v_bt}^{(\rm in)}$, and $\kappa_{\rm tot}=\kappa_{\rm ex} + \kappa_{\rm int}$.

Now, we assume that an input microwave is applied from the TL attached to the parametron.
We consider a continuous mode version of a coherent state:
\begin{eqnarray}
|\Psi_i\rangle = \mathcal{N}\exp\Big{[}\int_{-\infty}^{0}dr E_{\rm in}(-r) \tilde{b}_{r}^\dagger(0) \Big{]} |v\rangle,
\label{Psii_6_28_17}
\end{eqnarray}
with the overall vacuum state $|v\rangle$ and a normalization factor $\mathcal{N}$.
Considering that the input wave propagates in the positive-$r$ direction,
$E_{\rm in}(r)$ represents the input microwave at the initial moment as given by
\begin{eqnarray}
E_{\rm in}(r) = \left\{
\begin{array}{cl}
E e^{- i \omega_{\rm in} r/v_b} & (r>0) \\
0 & ({\rm otherwise}),
\end{array}
\right.
\label{E_9_29_20}
\end{eqnarray}
where $E$ and $\omega_{\rm in}$ are the amplitude and the angular frequency of the incoming microwave, respectively.
We assume that at the initial time the parametron is unexcited, and the input microwave has not arrived at the parametron yet.
Because $|\Psi_i\rangle$ is a coherent state, it is an eigenstate of the initial field operator $\tilde{b}_{r}(0)$.
We can obtain
\begin{eqnarray}
\tilde{b}_{v_bt}^{({\rm in})}(0) |\Psi_i\rangle = E_{\rm in}(v_bt) |\Psi_i\rangle 
= E e^{-i\omega_{\rm in}t}  |\Psi_i\rangle
\label{blvt_6_28_17}
\end{eqnarray}
using equations~(\ref{Psii_6_28_17}) and (\ref{E_9_29_20}).

We multiply equation~(\ref{br_2_18_21}) by $e^{i\omega_p t/2}$ with $r=+0$ and take the expectation value with respect to $|\Psi_i\rangle$ to obtain 
\begin{eqnarray}
\langle \tilde{B}_{+0}^{(\rm out)}(t) \rangle = \langle \tilde{B}_{v_bt}^{(\rm in)}(0) \rangle 
- i \sqrt\frac{\kappa_{\rm ex}}{v_b} \langle A(t) \rangle,
\label{in_out_6_28_17_2}
\end{eqnarray}
where 
\begin{eqnarray}
\tilde{B}_{+0}^{(\rm out)}(t) &=& \tilde{b}_{+0}^{(\rm out)}(t) e^{i\omega_p t/2},\nonumber\\
\tilde{B}_{v_bt}^{(\rm in)}(0) &=& \tilde{b}_{v_bt}^{(\rm in)}(0) e^{i\omega_p t/2},
\label{B_2_27_18_2}
\end{eqnarray}
and 
\begin{eqnarray}
A(t) = e^{i\omega_pt/2} a(t).
\end{eqnarray}
Equation~(\ref{blvt_6_28_17}) leads to 
\begin{eqnarray}
\langle \tilde{B}_{v_bt}^{(\rm in)}(0) \rangle = E e^{i(\omega_p /2-\omega_{\rm in})t}.
\label{Bin_10_29_20}
\end{eqnarray}

In this paper, we focus on the Fourier component of $\langle \tilde{B}_{+0}^{(\rm out)}(t) \rangle$ with a frequency of $-\omega_{\rm in}+\omega_p /2$, which is the same as the frequency of the input field, although our formalism can be used to obtain other frequency components of the reflected field.
We define the reflection coefficient as 
\begin{eqnarray}
\Gamma = \langle \tilde{B}_{+0}^{(\rm out)} \rangle[-\omega_{\rm in}+\omega_p /2]  / E,
\label{Gamma_2_19_21}
\end{eqnarray}
where 
$\langle \tilde{B}_{+0}^{(\rm out)} \rangle[-\omega_{\rm in}+\omega_p /2]$ is the Fourier component of $\langle \tilde{B}_{+0}^{(\rm out)} (t) \rangle$ with a frequency of $-\omega_{\rm in}+\omega_p /2$.
Equation (\ref{Gamma_2_19_21}) can be rewritten using equations~(\ref{in_out_6_28_17_2}) and (\ref{Bin_10_29_20}) as
\begin{eqnarray}
\Gamma = 1 - \frac{i}{E}\sqrt{\frac{\kappa_{\rm ex}}{v_b}} \langle A\rangle [-\omega_{\rm in} + \omega_p/2],
\label{Gamma_5_25_21}
\end{eqnarray}
where $ \langle A\rangle [-\omega_{\rm in} + \omega_p/2]$ is the Fourier component of $ \langle A(t)\rangle$ with a frequency of $-\omega_{\rm in}+\omega_p /2$.

\subsection{Representation of reflection coefficient with density matrix elements}
\label{Representation of reflection coefficient with density matrix elements}
We consider equations of motion of the system under consideration to calculate the reflection coefficient.
The time evolution of $\langle A\rangle$ is governed by
\begin{eqnarray}
\frac{d}{dt}\langle A\rangle = \Big{(} - i\Delta   - \frac{\kappa_{\rm tot}}{2} \Big{)} \langle A\rangle
+ i\chi \langle A^\dagger A^2 \rangle - 2i \beta \langle  A^\dagger \rangle
- i \sqrt{v_b\kappa_{\rm ex}} E e^{i(\omega_p/2-\omega_{\rm in})t},\nonumber\\
\label{Al_6_6_17}
\end{eqnarray}
where we used equations~(\ref{a_6_6_17_2}) and (\ref{blvt_6_28_17}) and omitted rapidly oscillating terms (rotating wave approximation). 
The approximation is valid when $\omega_p\gg \beta, \Delta, \chi$.
Effects of these rapidly oscillating terms on controls of a parametron were studied in Ref.~\cite{Masuda2020}.
Here, $\Delta$ is the detuning defined  by
\begin{eqnarray}
\Delta = \omega - \chi - \omega_p/2.
\end{eqnarray}
The master equation, which  leads to the same equations of motion, is represented as
\begin{eqnarray}
\frac{d\rho}{dt} = -\frac{i}{\hbar}[\mathcal{H}_{\rm sys}(t),\rho] + \mathcal{L}[\rho],
\label{drho_3_29_18}
\end{eqnarray}
with
\begin{eqnarray}
\mathcal{H}_{\rm sys} &=&  H_0 + \hbar\sqrt{v\kappa_{\rm ex}} [Ee^{-i(\omega_{\rm in}-\omega_p/2) t}A\da + Ee^{i(\omega_{\rm in}-\omega_p/2) t} A],
\label{H_9_29_20}\\
H_0 &=& \hbar \Delta A^\dagger A - \frac{\hbar\chi}{2}A^{\dagger 2}A^2 + \hbar\beta(A^2 + A^{\dagger 2})\label{H_RWA_3_18_21}\\
\mathcal{L}[\rho] &= &\frac{\kappa_{\rm tot}}{2} \Big{(} [A\rho,A\da] + [A,\rho A\da] \Big{)},
\end{eqnarray}
where $\rho$ is the density operator. 

In order to derive an analytic formula of the reflection coefficient, we use energy eigenstates $|\phi_m\rangle$ of $H_0$ in equation~(\ref{H_RWA_3_18_21}) to rewrite $\langle A \rangle$ as
\begin{eqnarray}
\langle A \rangle &=& {\rm Tr}[A\rho] = \sum_{m} \langle \phi_m | A\rho |\phi_m \rangle = \sum_{mn} X_{mn}\rho_{nm},
\label{A_3_29_21}
\end{eqnarray}
where $\rho_{nm} = \langle \phi_n | \rho | \phi_m \rangle$, and 
$X_{mn} = \langle \phi_m | A | \phi_n \rangle$.
Using equations~(\ref{Gamma_5_25_21}) and (\ref{A_3_29_21}), the reflection coefficient can be represented as 
\begin{eqnarray}
\Gamma  = 1 - \frac{i}{E} \sqrt\frac{\kappa_{\rm ex}}{v_b} \sum_{mn} X_{mn}\rho^{\rm (F)}_{nm}[-\omega_{\rm in}+\omega_p /2],
\label{Gamma_2_19_21_2}
\end{eqnarray}
where $\rho^{\rm (F)}_{nm}[-\omega_{\rm in}+\omega_p /2]$ is the Fourier component of $\rho_{nm}$ at a frequency of $-\omega_{\rm in}+\omega_p /2$.
The term proportional to $\rho^{\rm (F)}_{nm}[-\omega_{\rm in}+\omega_p /2] $ in equation~(\ref{Gamma_2_19_21_2})
is the contribution to the reflection coefficient from the transition from $|\phi_m\rangle$ to $|\phi_n\rangle$.

The equation of motion of the density matrix element is written as
\begin{eqnarray}
\dot\rho_{mn} &=& i(-\omega_m + \omega_n) \rho_{mn} 
- i\Omega \sum_k (X_{mk} \rho_{kn} - X_{kn} \rho_{mk} ) e^{i(\omega_{\rm in} - \omega_p/2)t}\nonumber\\
&& - i\Omega \sum_k (X_{km}^\ast \rho_{kn} - X_{nk}^\ast \rho_{mk} ) e^{-i(\omega_{\rm in} - \omega_p/2)t}\nonumber\\
&& + \kappa_{\rm tot} \sum_{kl} X_{mk} X_{nl}^\ast \rho_{kl} 
-\frac{\kappa_{\rm tot} }{2}  \sum_{k} (Y_{mk} \rho_{kn} + Y_{kn} \rho_{mk} ),
\label{drho_3_18_21}
\end{eqnarray}
where $\omega_m$ is an eigenvalue of $H_0/\hbar$; $Y_{mn}=\langle \phi_m|A^\dagger A |\phi_n\rangle$; and
$\Omega$ is defined by 
\begin{eqnarray}
\Omega=\sqrt{v_b\kappa_{\rm ex}}E.
\end{eqnarray}
The Fourier transform of equation (\ref{drho_3_18_21}) with a  frequency of $-\omega_{\rm in} + \omega_p/2~ ( = -\tilde\omega_{\rm in})$ leads to
\begin{eqnarray}
0 &=& i(\omega_{\rm in} - \omega_p/2 -\omega_m + \omega_n)\rho^{\rm (F)}_{mn}[-\tilde\omega_{\rm in}]
- i\Omega \sum_k (X_{mk} \rho^{\rm (F)}_{kn}[-2\tilde\omega_{\rm in}] - X_{kn} \rho^{\rm (F)}_{mk}[-2\tilde\omega_{\rm in}] )\nonumber\\
&& - i\Omega \sum_k (X_{km}^\ast \rho^{\rm (F)}_{kn}[0] - X_{nk}^\ast \rho^{\rm (F)}_{mk}[0] ) 
+ \kappa_{\rm tot} \sum_{kl} X_{mk} X_{nl}^\ast \rho^{\rm (F)}_{kl} [-\tilde\omega_{\rm in}]
\nonumber\\
&&  -\frac{\kappa_{\rm tot} }{2}  \sum_{k} (Y_{mk} \rho^{\rm (F)}_{kn}[-\tilde\omega_{\rm in}] + Y_{kn} \rho^{\rm (F)}_{mk}[-\tilde\omega_{\rm in}] ).
\label{rho_F_v1_2_20_21}
\end{eqnarray}
These equations are used to obtain the density matrix elements in equation~(\ref{Gamma_2_19_21_2}) as shown in the following section.

\section{Weak input field limit}
\label{Weak input field limit}
In principle, the reflection coefficient in equation (\ref{Gamma_5_25_21}) can be obtained using the density matrix which can be calculated by integrating the master equation~(\ref{drho_3_29_18}).
 However, it is time consuming to integrate the master equation for sufficiently long time for a parametron.
In this section, we present an alternative method providing an approximate reflection coefficient in the weak input field regime, where the diagonal elements of the density matrix $\rho^{\rm (F)}_{mm}[0]$ are approximately the same as those for the stationary state without input field.
The method does not require to integrate the master equation. The effect of the finite input field can be also taken into account, which will be discussed in \ref{Effect of input field}.

We consider the case that the input field is nearly resonant with the transition from $|\phi_m\rangle$ to $|\phi_n\rangle$, that is, $\omega_{\rm in} - \omega_p/2 + \omega_m - \omega_n \simeq 0$.
We assume that the off-diagonal element of the density matrix $\rho^{\rm (F)}_{kl}[-\tilde\omega_{\rm in}]$ is not vanishing only for $(k,l)=(n,m)$, and non-resonant elements such as $\rho^{\rm (F)}_{kl}[-2\tilde\omega_{\rm in}]$ are vanishing.
We also assume that the diagonal elements are the same as the stationary state without input field because the input field is sufficiently weak.
Then, $\rho^{\rm (F)}_{nm}[-\tilde\omega_{\rm in}]$ can be obtained using equation~(\ref{rho_F_v1_2_20_21}) as
\begin{eqnarray}
{\rho^{\rm (F)}}_{nm}[-\tilde\omega_{\rm in}] = \frac{i\Omega X_{mn}^\ast (\rho^{\rm (F)}_{mm}[0] - \rho^{\rm (F)}_{nn}[0]) }{i\Delta_{nm}- \frac{\kappa_{\rm ex}+\kappa_{\rm int}}{2}(Y_{mm}+Y_{nn})}.
\label{rhomn_2_19_21}
\end{eqnarray}
where $\Delta_{nm} = \omega_{\rm in} - \omega_p/2 - \omega_n + \omega_m$.
We used $X_{mm}=0$ in equation~(\ref{rhomn_2_19_21}), which is valid because $|\phi_m\rangle$ has even or odd parity, that is, $|\phi_m\rangle$ is a superposition of even-photon-number states or odd-photon-number states.
Using equation~(\ref{rhomn_2_19_21}) in equation~(\ref{Gamma_2_19_21_2}), the reflection coefficient is approximately represented by the following simple form,
\begin{eqnarray}
\Gamma = 1 + \sum_{mn}\xi_{mn}
\label{Gamma_3_30_21}
\end{eqnarray}
with 
\begin{eqnarray}
\xi_{mn}  =  \frac{\kappa_{\rm ex} |X_{mn}|^2 (\rho^{\rm (F)}_{mm}[0] - \rho^{\rm (F)}_{nn}[0]) }{i\Delta_{nm}- \frac{\kappa_{\rm ex}+\kappa_{\rm int}}{2}(Y_{mm}+Y_{nn})}.
\label{Gamma_2_19_21_3}
\end{eqnarray}
Note that the diagonal elements of the density matrix in equation (\ref{Gamma_2_19_21_3}) are for the stationary state of the case without input field. 
When either of the resonant energy levels is occupied, the amplitude of the reflection coefficient changes from unity, and thus a dip or peak in the reflection coefficient appears.


Around a dip or peak corresponding to the transition from $|\phi_m\rangle$ to $|\phi_n\rangle$, $\Gamma$ is approximated by $\Gamma_{mn}$ defined by
\begin{eqnarray}
\Gamma_{mn} = 1 + \xi_{mn}
= 1 + \frac{\tilde\kappa^{(mn)}_{\rm ex}}{i\Delta_{nm} - (\tilde\kappa^{(mn)}_{\rm ex}+\tilde\kappa^{(mn)}_{\rm int})}
\label{Gamma_mn_2_20_21}
\end{eqnarray}
with 
\begin{eqnarray}
\tilde\kappa^{(mn)}_{\rm ex} &=& \kappa_{\rm ex} |X_{mn}|^2 (\rho^{\rm (F)}_{mm}[0] - \rho^{\rm (F)}_{nn}[0]),\nonumber\\
\tilde\kappa^{(mn)}_{\rm int} &=& (\kappa_{\rm ex}+\kappa_{\rm int})(Y_{mm}+Y_{nn}) - \kappa_{\rm ex} |X_{mn}|^2 (\rho^{\rm (F)}_{mm}[0] - \rho^{\rm (F)}_{nn}[0]).
\label{nominal_3_18_21}
\end{eqnarray}
On the other hand, the reflection coefficient of a linear resonator is represented as~\cite{Walls2008} 
\begin{eqnarray}
\Gamma_{\rm r} = 1 + \frac{\kappa^{(r)}_{\rm ex}}{i\Delta_{\rm r} - (\kappa^{(r)}_{\rm ex}+\kappa^{(r)}_{\rm int})},
\label{Gamma_r_2_20_21}
\end{eqnarray}
where $\kappa^{(r)}_{\rm ex}$ and $\kappa^{(r)}_{\rm int}$ are the external and the internal decay rates; $\Delta_{\rm r} =\omega_{\rm in} - \omega_0$; and $\omega_0$ is the angular resonance frequency of the resonator.
Thus, $\tilde\kappa^{(mn)}_{\rm ex}$ and $\tilde\kappa^{(mn)}_{\rm int}$ can be interpreted as the nominal external and internal decay rates, respectively.

The internal and the external decay rates of a linear resonator can be obtained via a fitting of the measured reflection coefficient to the analytic form of equation~(\ref{Gamma_r_2_20_21}). The nominal decay rates of the parametrons can be obtained in the same manner with equation~(\ref{nominal_3_18_21}), and can provide information of the measured parametron such as $|X_{mn}|^2 (\rho^{\rm (F)}_{mm}[0] - \rho^{\rm (F)}_{nn}[0])$ and $Y_{mm}+Y_{nn}$. Note that $\kappa_{\rm ex}$ and $\kappa_{\rm int}$ in equation~(\ref{nominal_3_18_21}) can be obtained via measurements of the parametron without pump field.

The measurement of $\tilde\kappa_{\rm ex}^{mn}$ is rather useful in the strong pump regime, where $\beta$ is sufficiently larger than $\chi$ and $\Delta$.
It is known that the stationary state is the maximally mixed state of the two highest levels in the strong pump regime \cite{Goto2016}.
For example, in the case of $\Delta\le 0$, we have $\rho_{00}^{(F)}[0]=\rho_{11}^{(F)}[0]=1/2$ and $\rho_{mm}^{(F)}[0]=0$ for $m\ge 2$ for sufficiently large $\beta$ as shown later.
Then, equation (\ref{nominal_3_18_21}) gives the amplitude of the coupling coefficient $|X_{mn}|$ between either of the two highest levels and a lower level as $|X_{mn}| = \sqrt{2\tilde\kappa_{\rm ex}^{mn}/\kappa_{\rm ex}}$, where $m=0,1$ and $n\ge 2$. Therefore, the amplitude of the coupling coefficients can be directly extracted by the measurement of $\tilde\kappa_{\rm ex}^{mn}$. 
Recently, a fast gate operation of a parametron utilizing energy levels outside of the qubit space was proposed \cite{Kanao2021b}.
It is useful to experimentally extract $|X_{mn}|$ for tailoring a control field in such protocols to improve the gate fidelity.

The reflection coefficient can be calculated also by a straightforward but time consuming manner of
integrating the master equation~(\ref{drho_3_29_18}).
In \ref{Comparison with results with master equation}, we compare the results of the two methods to numerically verify the approximate method.

\section{Numerical results}
\begin{figure}[h!]
\begin{center}
\includegraphics[width=14cm]{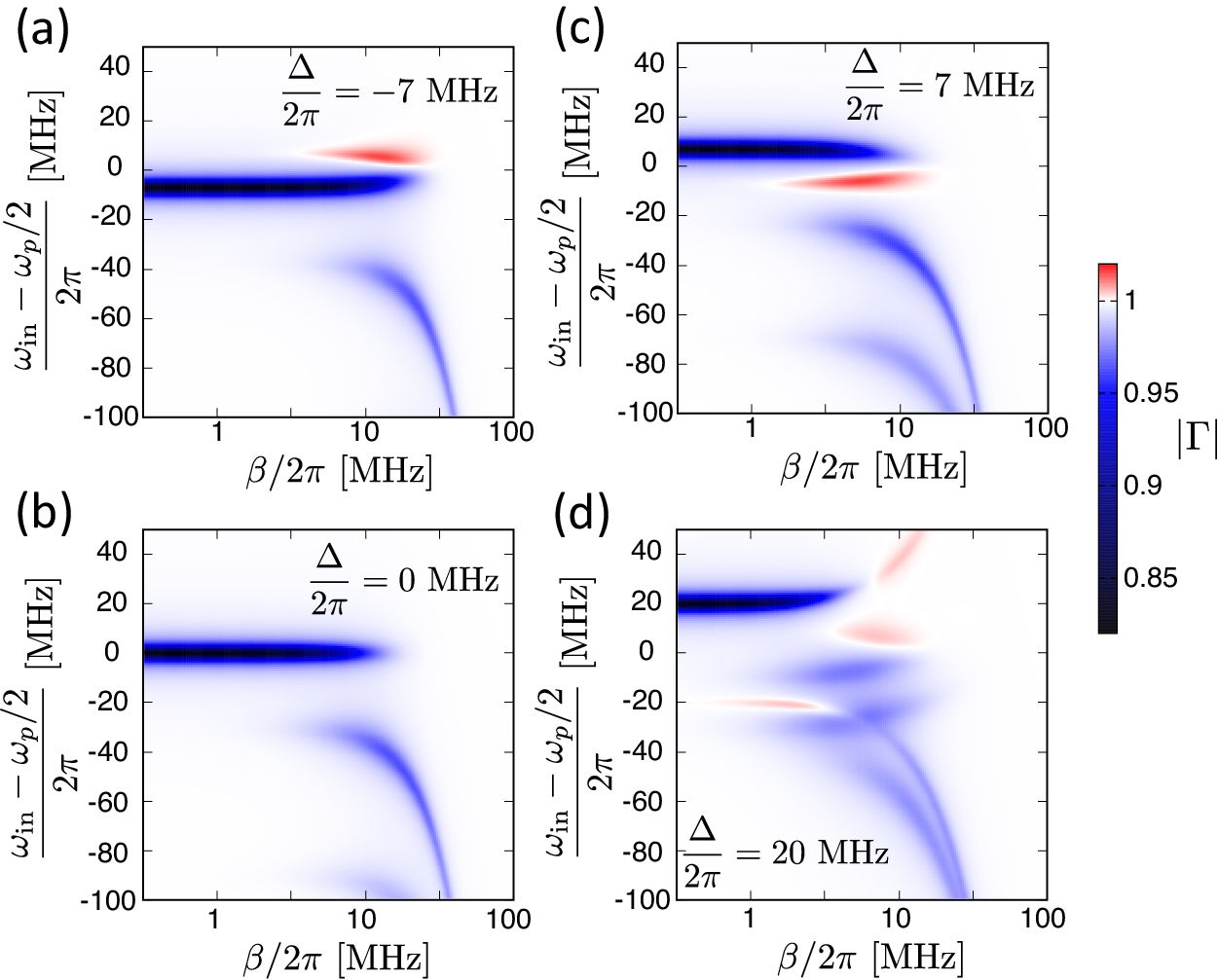}
\end{center}
\caption{
Amplitude of the reflection coefficient in equation~(\ref{Gamma_3_30_21}) in the weak input field limit as a function of $\omega_{\rm in}$ and $\beta$ for $\Delta/2\pi=-7$~MHz (a), 
0~MHz (b), 7~MHz (c) and 20~MHz (d).
The used parameter set is: $\chi/2\pi = 30$~MHz, $\kappa_{\rm ex}/2\pi=0.4$~MHz and $\kappa_{\rm int}/2\pi=4$~MHz.}
\label{DeltaE_woline_3_16_21}
\end{figure}

Figure~\ref{DeltaE_woline_3_16_21} shows the amplitude of the reflection coefficient in equation~(\ref{Gamma_3_30_21}) as a function of the angular frequency of the incoming microwave $\omega_{\rm in}$ and the amplitude of the pump field~$\beta$ for various values of the detuning $\Delta$.
Here, $\rho^{\rm (F)}_{mm}[0]$ and $\rho^{\rm (F)}_{nn}[0]$ in equation~(\ref{Gamma_2_19_21_3}) were numerically calculated using equation~(\ref{drho_3_18_21}).
The used parameter set is: $\chi/2\pi = 30$~MHz, $\kappa_{\rm ex}/2\pi=0.4$~MHz and $\kappa_{\rm int}/2\pi=4$~MHz, and is typical for superconducting parametrons~\cite{Wang2019,Yamaji2021}. 
When $\beta\simeq 0$, there is only a single dip of $|\Gamma|$ at $\omega_{\rm in}-\omega_p/2 = \Delta$ corresponding to the transition from Fock state $|0\rangle$ to Fock state $|1\rangle$ as seen in figure~\ref{DeltaE_woline_3_16_21}.
On the other hand, the spectra show the interesting behaviors as $\beta$ increases: the dip corresponding to the transition $|0\rangle \rightarrow |1\rangle$ disappears while other peaks and dips appear; 
the frequencies corresponding to some of the peaks and dips increase with $\beta$, while the others decrease; a dip (peak) changes to a peak (dip) in figure~\ref{DeltaE_woline_3_16_21}(d).
And, the pattern of the spectrum depends on the detuning.
In the following, these behaviors of $|\Gamma|$ are explained.

\begin{figure}[h!]
\begin{center}
\includegraphics[width=12cm]{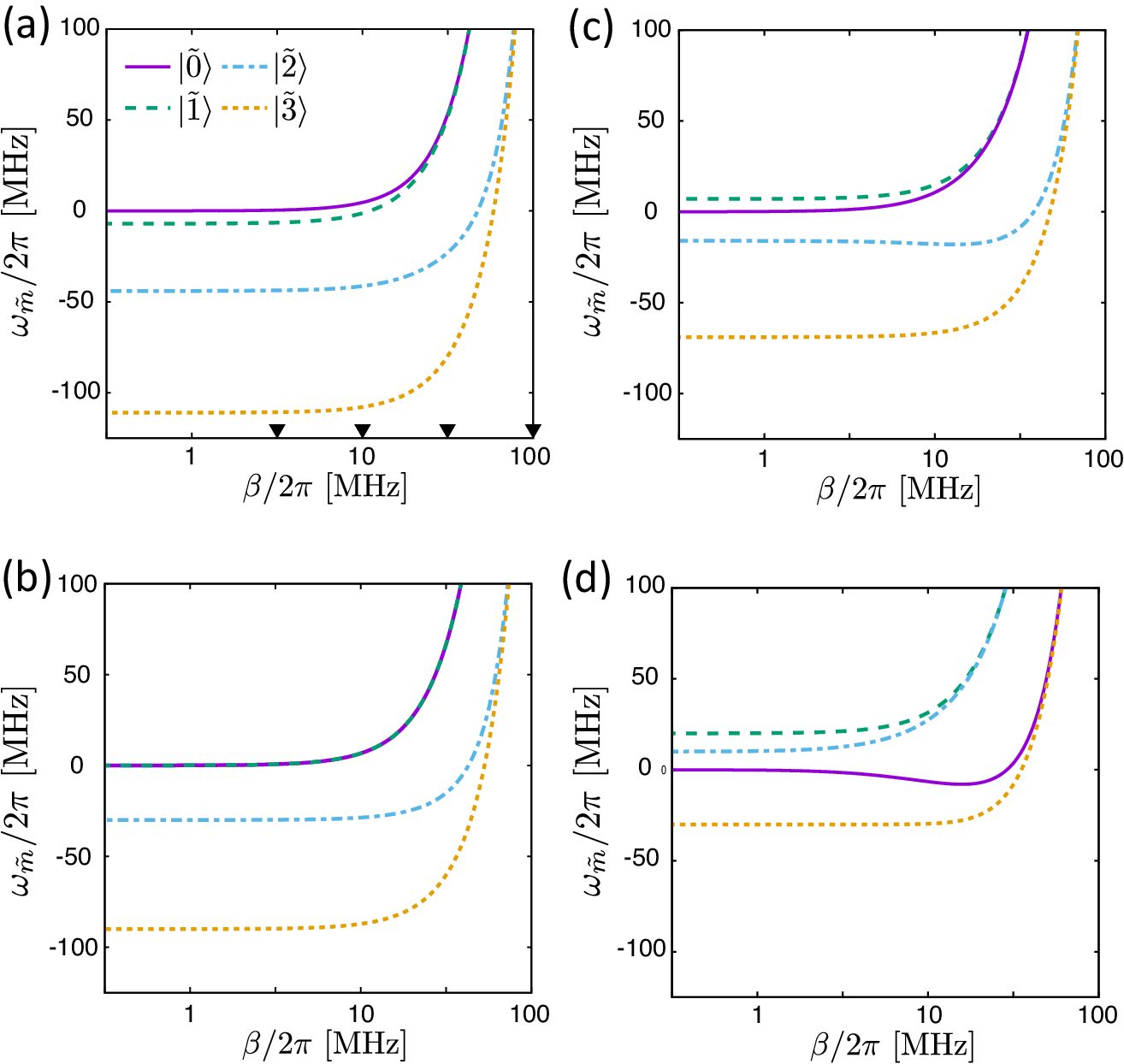}
\end{center}
\caption{
Energy diagram of $H_0$ for $\Delta/2\pi=-7$~MHz (a), 0~MHz (b), 7~MHz (c) and 20~MHz (d).
Only four levels are displayed.
The other used parameters are the same as figure~\ref{DeltaE_woline_3_16_21}.
Wigner function of the stationary state is shown in figure~\ref{Wig_3_17_21} for the values of $\beta$ indicated by the triangles in panel (a).}
\label{eng_3_16_21}
\end{figure}

\begin{figure}[h!]
\begin{center}
\includegraphics[width=14cm]{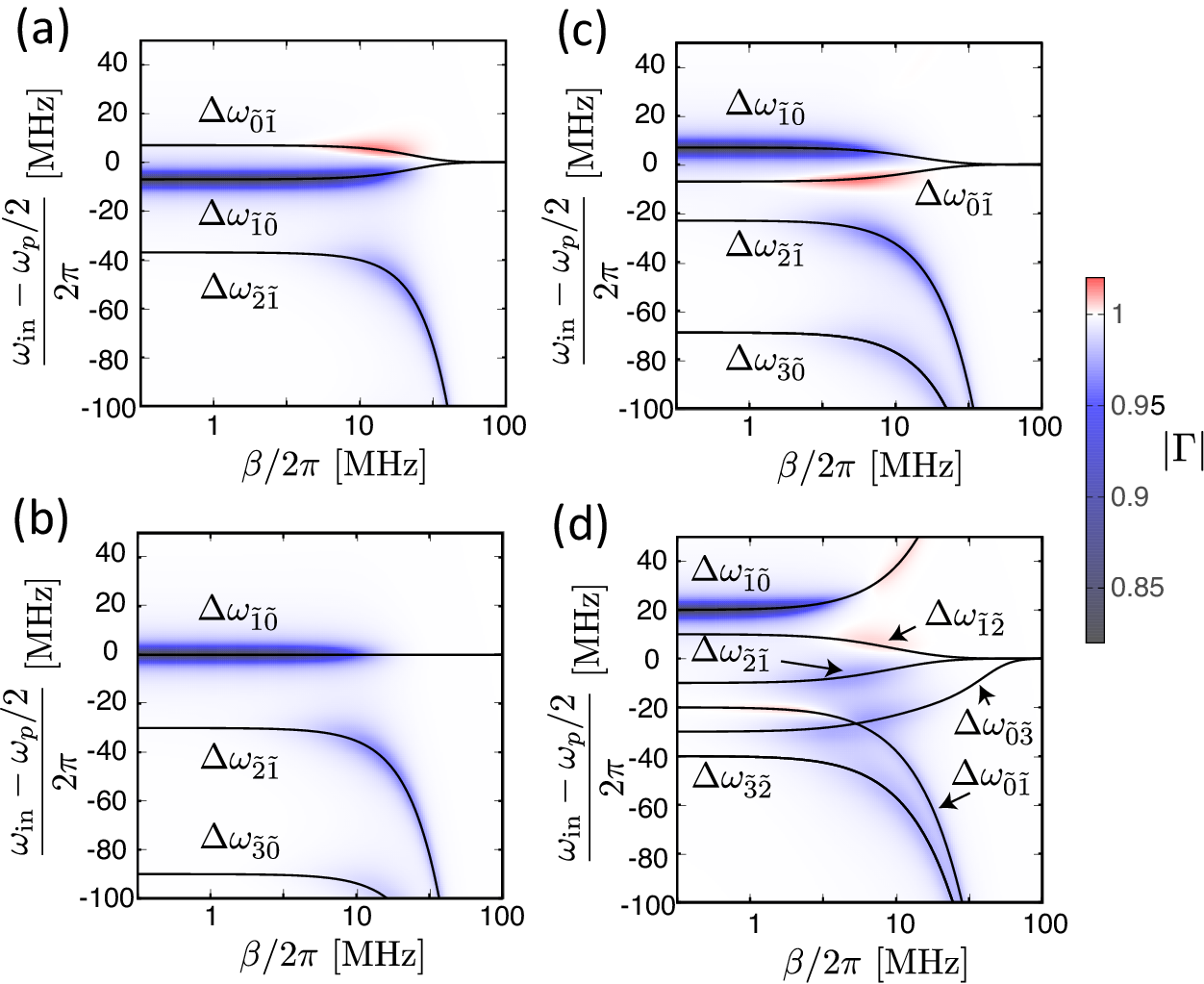}
\end{center}
\caption{
Spectra in figure~\ref{DeltaE_woline_3_16_21} compared with curves representing the energy differences, $\Delta\omega_{\tilde{n}\tilde{m}} = \omega_{\tilde{n}} - \omega_{\tilde{m}}$, corresponding to the transition from $|\tilde{m}\rangle$ to $|\tilde{n}\rangle$.
(The color of $|\Gamma|$ is chosen to be lighter than figure~\ref{DeltaE_woline_3_16_21} to make the curves clear.)
Only curves for relevant transitions are shown.
The used parameters are the same as figure~\ref{DeltaE_woline_3_16_21}.}
\label{DeltaE2_3_16_21}
\end{figure}

Each dip and peak of $|\Gamma|$ corresponds to a transition between eigenstates of $H_0$ in equation~(\ref{H_RWA_3_18_21}).
Therefore, it is useful to examine the eigenvalues of $H_0$ shown in figure~\ref{eng_3_16_21}.
Each eigenstate of $H_0$ denoted by $|\tilde{m}\rangle$ coincides with Fock state $|m\rangle$ when $\beta=0$.
We denote the eigenenergy as $\omega_{\tilde{m}}$.
Figure~\ref{DeltaE2_3_16_21} shows the same spectra as figure~\ref{DeltaE_woline_3_16_21} with the curves representing energy differences, $\Delta\omega_{\tilde{n}\tilde{m}} = \omega_{\tilde{n}} - \omega_{\tilde{m}}$, corresponding to the transition, $|\tilde{m}\rangle\rightarrow |\tilde{n}\rangle$.
The dips and peaks match to the curves for the energy differences.
Thus, the spectra reflect the energy level structure of $H_0$.
Therefore, the spectra can provide information on the energy level structure of the parametron.
As seen in figure~\ref{eng_3_16_21}, the order of levels, $|\tilde{m}\rangle$, depends on the detuning \cite{Zhang2017}.
Thus, the pattern of the spectrum also changes depending on the detuning.
For example, the order of the distinct dip and peak around $\omega_{\rm in}=\omega_{p}/2$ are opposite in figures~\ref{DeltaE2_3_16_21}(a) and \ref{DeltaE2_3_16_21}(c). It reflects the difference in the order of $|\tilde{0}\rangle$ and $|\tilde{1}\rangle$ observed in figures~\ref{eng_3_16_21}(a) and \ref{eng_3_16_21}(c).

As seen from equation~(\ref{Gamma_2_19_21_3}),  the finite population difference $(|\rho^{\rm (F)}_{\tilde{m}\tilde{m}}[0] -\rho^{\rm (F)}_{\tilde{n}\tilde{n}}[0]|>0)$ and the finite coupling coefficient $(|X_{\tilde{m}\tilde{n}}|>0)$ are required for the corresponding dip or peak of $|\Gamma|$ to be visible.
This explains the appearance and the disappearance of the dips and the peaks in figure~\ref{DeltaE_woline_3_16_21}.
In the following, we first look into the population of relevant levels, and then examine relevant coupling coefficients.

\begin{figure}[h!]
\begin{center}
\includegraphics[width=12cm]{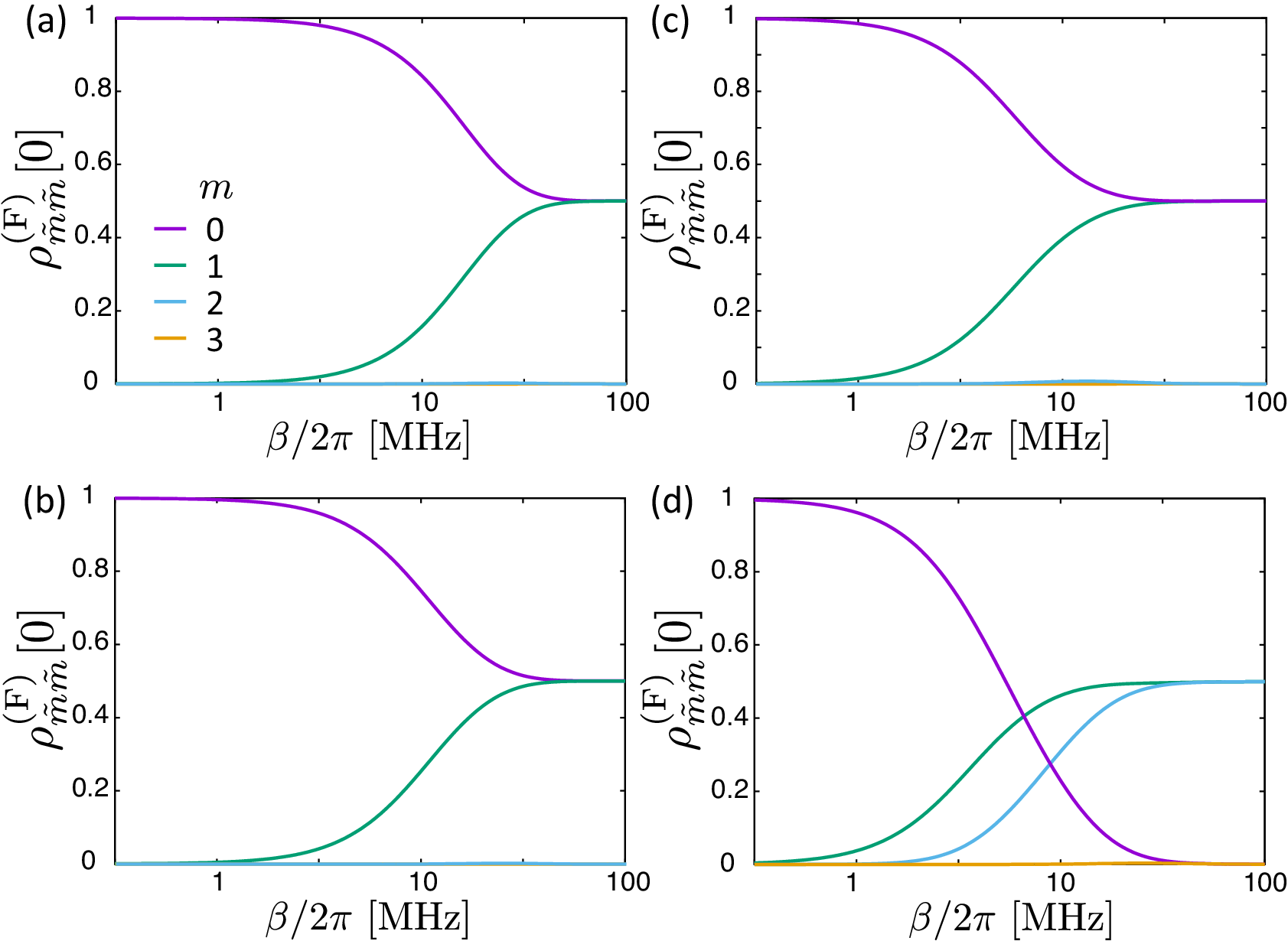}
\end{center}
\caption{
Population of each level, $\rho^{\rm (F)}_{\tilde{m}\tilde{m}}[0]$, for the stationary state in the weak input field limit for $\Delta/2\pi=-7$~MHz (a), 0~MHz (b), 7~MHz (c) and 20~MHz (d).
$\rho^{\rm (F)}_{\tilde{2}\tilde{2}}[0]$ is approximately zero in (a)--(c), while 
 $\rho^{\rm (F)}_{\tilde{3}\tilde{3}}[0]$ is approximately zero in (a)--(d).
The other used parameters are the same as figure~\ref{DeltaE_woline_3_16_21}.
The stationary state approaches to the maximally mixed state of the highest and the second highest levels as $\beta$ increases.}
\label{rho_com_3_17_21}
\end{figure}

Figure~\ref{rho_com_3_17_21} shows the population of each level, $\rho^{\rm (F)}_{\tilde{m}\tilde{m}}[0]$, for the stationary state.
The stationary state for $\beta=0$ is $|0\rangle$ because of the decay.
Thus, $\rho^{\rm (F)}_{\tilde{0} \tilde{0}}[0]\simeq 1$ and $\rho^{\rm (F)}_{\tilde{m}\tilde{m}}[0]\simeq 0$ for $m\ne 0$ for $\beta\simeq 0$.
When $\beta\gg |\Delta|,\chi$, the highest and the second highest levels become the even or the odd cat states.
\footnote{For example, the highest and the second highest levels become the even and the odd cat states, respectively, for $\Delta<0$.}
The even and odd cat states are represented as $(|\alpha\rangle + |-\alpha\rangle)/\sqrt{2}$ and $(|\alpha\rangle - |-\alpha\rangle)/\sqrt{2}$, respectively, where $\alpha\simeq\sqrt{2\beta/\chi}$.
Because of the decay, the stationary state becomes the maximally mixed state of $|\alpha\rangle$ and $|-\alpha\rangle$ for $\beta\gg |\Delta|,\chi$ \cite{Puri2017b}, which is the same as the maximally mixed state of the highest and the second highest levels. 
This is consistent with the obtained results in figure~\ref{rho_com_3_17_21}, that is, for large $\beta$, $\rho^{\rm (F)}_{\tilde{0} \tilde{0}}[0]=\rho^{\rm (F)}_{\tilde{1} \tilde{1}}[0]=1/2$ for $\Delta/2\pi=-7$, 0, 7~MHz and $\rho^{\rm (F)}_{\tilde{1} \tilde{1}}[0]=\rho^{\rm (F)}_{\tilde{2} \tilde{2}}[0]=1/2$ for $\Delta/2\pi=20$~MHz.
Note that $|\tilde{1}\rangle$ and $|\tilde{2}\rangle$ are the highest and the second highest levels for $\Delta/2\pi=20$~MHz, respectively.
The dip corresponding to $|\tilde{0}\rangle \rightarrow |\tilde{1}\rangle$ disappears for $\Delta/2\pi=-7$, 0, 7~MHz as $\beta$ increases, because $\rho^{\rm (F)}_{\tilde{0}\tilde{0}}[0] -\rho^{\rm (F)}_{\tilde{1}\tilde{1}}[0]$ vanishes.
 
\begin{figure}[h!]
\begin{center}
\includegraphics[width=12cm]{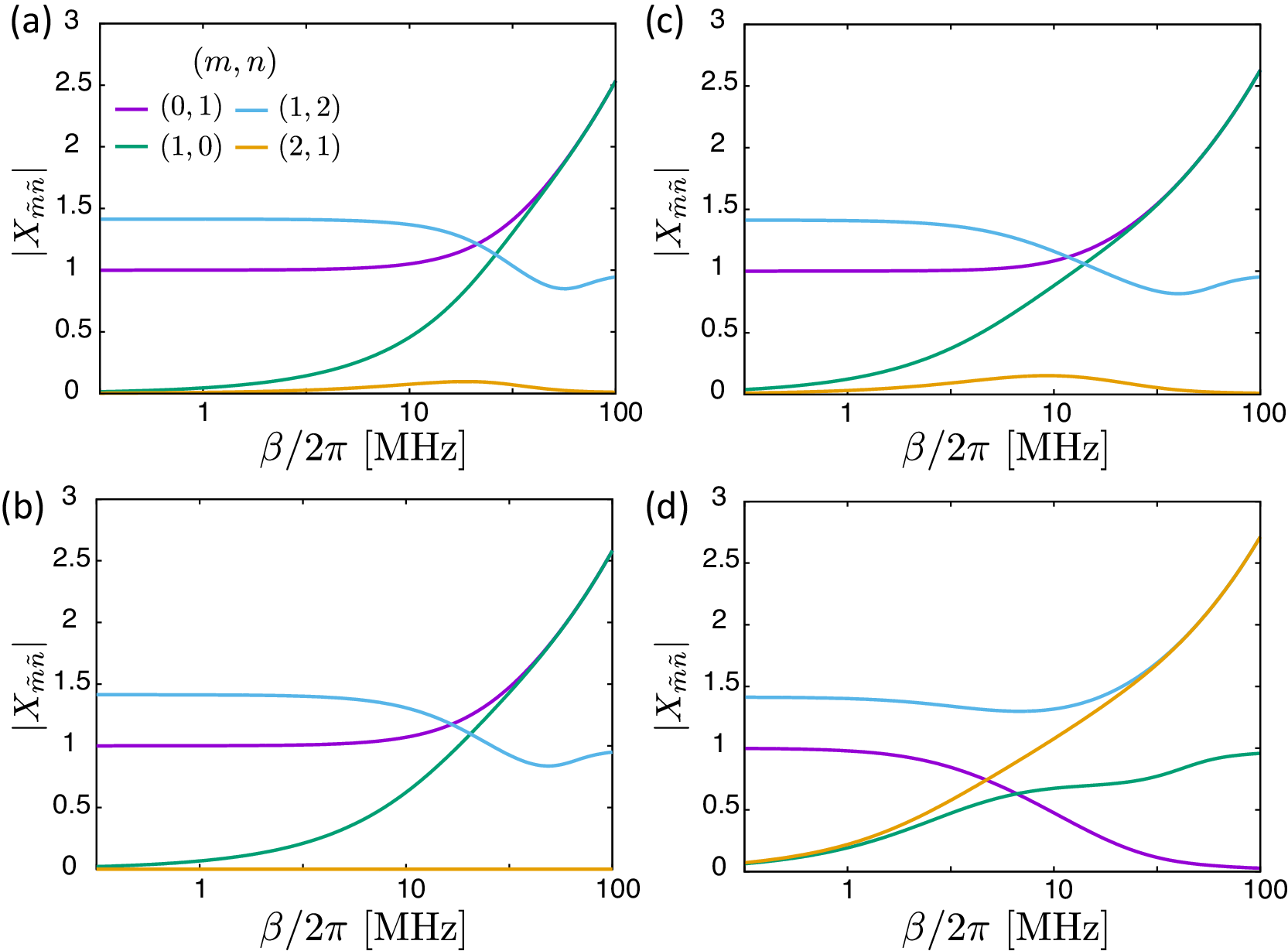}
\end{center}
\caption{
Amplitude of the coupling coefficient $X_{\tilde{m}\tilde{n}}$ for the transitions $|\tilde{0}\rangle\rightarrow |\tilde{1}\rangle$, $|\tilde{1}\rangle\rightarrow |\tilde{0}\rangle$, 
$|\tilde{1}\rangle\rightarrow |\tilde{2}\rangle$ and $|\tilde{2}\rangle\rightarrow |\tilde{1}\rangle$.
Panels (a)--(d) are for $\Delta/2\pi=-7$~MHz (a), 0~MHz (b), 7~MHz (c) and 20~MHz (d).
The other used parameters are the same as figure~\ref{DeltaE_woline_3_16_21}.}
\label{Xs_com_3_18_21}
\end{figure}

Coupling coefficients are shown for relevant transitions in figure~\ref{Xs_com_3_18_21}.
It is seen that $|X_{\tilde{0}\tilde{1}}|$, $|X_{\tilde{1}\tilde{0}}|$ for $\Delta/2\pi=-7$, 0, 7~MHz and 
$|X_{\tilde{1}\tilde{2}}|$, $|X_{\tilde{2}\tilde{1}}|$ for $\Delta/2\pi=20$~MHz increase rapidly with respect to $\beta$ when $\beta$ is sufficiently large ($\beta/2\pi > 10$~MHz).
This is because that the highest and the second highest levels are superpositions of $|\alpha\rangle$ and $|-\alpha\rangle$ when $\beta$ is sufficiently large, and $|X_{\tilde{m}\tilde{n}}|$ between these levels is approximately $\alpha (\simeq \sqrt{2\beta/\chi})$.
The profile of $|X_{\tilde{m}\tilde{n}}|$ for $\Delta/2\pi=20$~MHz is different from that for $\Delta/2\pi=-7$, 0, 7~MHz due to the difference in the order of the levels as represented in figure~\ref{eng_3_16_21}.
Note that $|X_{\tilde{1}\tilde{0}}|$ corresponding to $|\tilde{1}\rangle \rightarrow |\tilde{0}\rangle$ increases with respect to $\beta$.
This transition gives rise to a peak of $|\Gamma|$ for $\Delta/2\pi=-7$ and 7~MHz.
\begin{figure}[h!]
\begin{center}
\includegraphics[width=12cm]{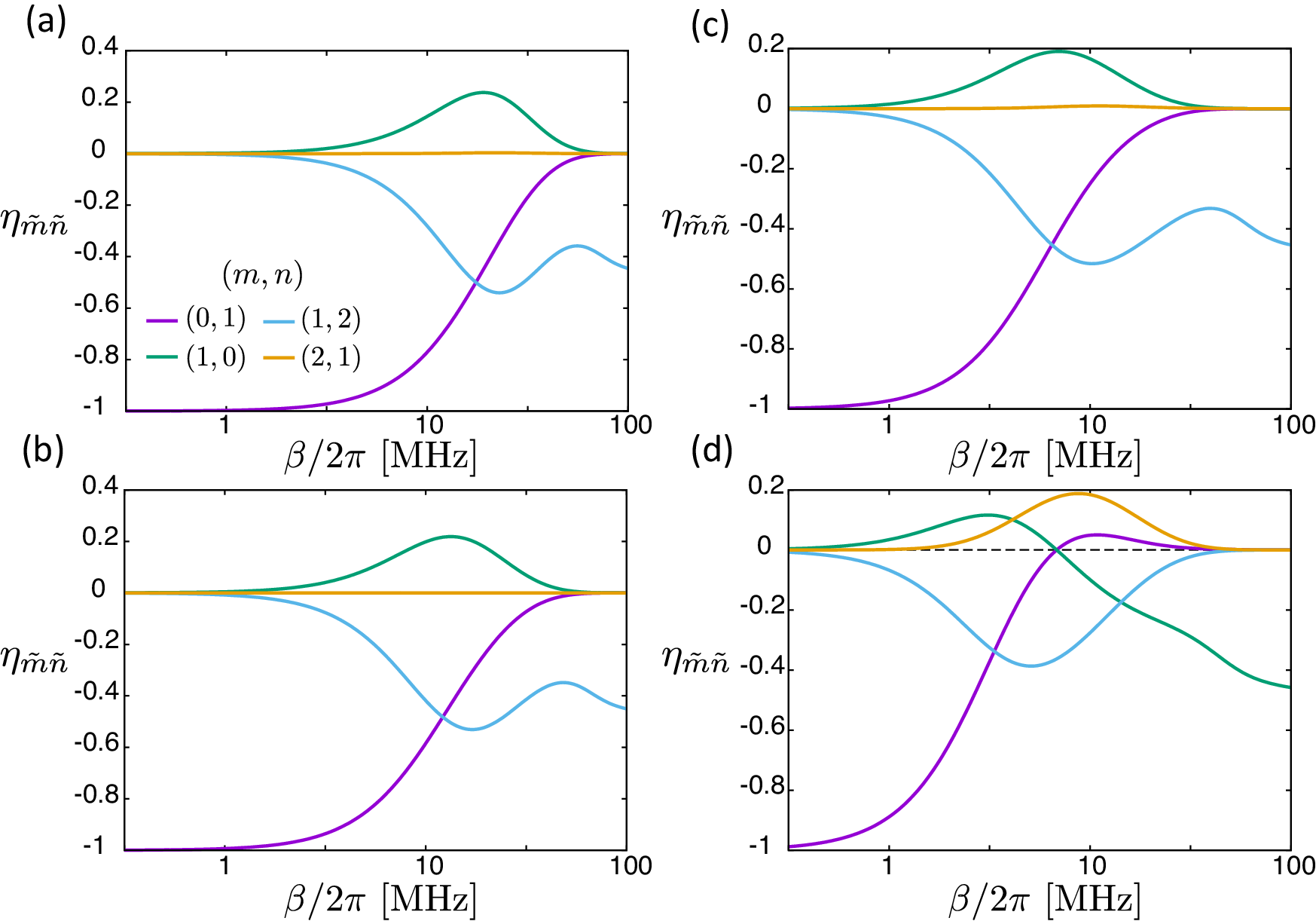}
\end{center}
\caption{
$\eta_{\tilde{m}\tilde{n}}=-|X_{\tilde{m}\tilde{n}}|^2(\rho^{\rm (F)}_{\tilde{m}\tilde{m}}[0]-\rho^{\rm (F)}_{\tilde{n}\tilde{n}}[0])$ is shown as a function of $\beta$ for $\Delta/2\pi=-7$~MHz (a), 0~MHz (b), 7~MHz (c) and 20~MHz (d). 
Positive and negative $\eta_{\tilde{m}\tilde{n}}$ correspond to a peak and a dip of $|\Gamma|$ for transition, $|\tilde{m}\rangle\rightarrow|\tilde{n}\rangle$, respectively.
The other used parameters are the same as figure~\ref{DeltaE_woline_3_16_21}.
}
\label{Xrho_3_18_21}
\end{figure}
 
It is useful to examine $\eta_{\tilde{m}\tilde{n}}=-|X_{\tilde{m}\tilde{n}}|^2(\rho^{\rm (F)}_{\tilde{m}\tilde{m}}[0]-\rho^{\rm (F)}_{\tilde{n}\tilde{n}}[0])$ in equation~(\ref{Gamma_2_19_21_3}) to explain the appearance and disappearance of dips and peaks of $|\Gamma|$. 
Note that $Y_{\tilde{m}\tilde{m}}$ and $Y_{\tilde{n}\tilde{n}}$ are always positive, and $\Delta_{\tilde{n}\tilde{m}}$ is zero at the resonance in equation~(\ref{Gamma_2_19_21_3}).
If $\eta_{\tilde{m}\tilde{n}}$ is positive, a peak corresponding to the transition $|\tilde{m}\rangle \rightarrow |\tilde{n}\rangle$ shows up. If negative, a dip appears. 
Figure~\ref{Xrho_3_18_21} shows $\eta_{\tilde{m}\tilde{n}}$ as a function of $\beta$ for four sets of $(\tilde{m},\tilde{n})$.
The results for $\Delta/2\pi=-7, 0, 7$~MHz indicate: 1. the dip for the transition $|\tilde{0}\rangle \rightarrow |\tilde{1}\rangle$ vanishes as $\beta$ increases; 2. the dip for $|\tilde{1}\rangle \rightarrow |\tilde{2}\rangle$ and the peak for $|\tilde{1}\rangle \rightarrow |\tilde{0}\rangle$ appear as $\beta$ increases, although the peak vanishes when $\beta$ is further increased; 3. transition $|\tilde{2}\rangle\rightarrow |\tilde{1}\rangle$ is hardly seen. This is because that the coupling coefficient, $X_{\tilde{2}\tilde{1}}$, is approximately zero as shown in figures~\ref{Xs_com_3_18_21}(a)$-$(c). 
The results for $\Delta/2\pi=20$~MHz indicate: 1. the dip for $|\tilde{1}\rangle \rightarrow |\tilde{2}\rangle$ and the peak for $|\tilde{2}\rangle \rightarrow |\tilde{1}\rangle$ appear for intermediate value of $\beta$; 2. the transition $|\tilde{0}\rangle \rightarrow |\tilde{1}\rangle$ gives rise to a dip of $|\Gamma|$ for relatively small $\beta$ and a peak for $\beta/2\pi \simeq 10$~MHz, and then vanishes as $\beta$ is further increased;
3. the  transition $|\tilde{1}\rangle \rightarrow |\tilde{0}\rangle$ gives rise to a peak for relatively small $\beta$ and a dip for relatively large $\beta$.  The change of the sign of $\eta_{\tilde{m}\tilde{n}}$ corresponding to the transitons $|\tilde{1}\rangle \leftrightarrow |\tilde{0}\rangle$
 comes from the crossing of the populations of $|\tilde{0}\rangle$ and $|\tilde{1}\rangle$ observed in figure~\ref{rho_com_3_17_21}(d).
Thus, these results explain the profile of the spectrum in figure~\ref{DeltaE_woline_3_16_21}.

\begin{figure}[h!]
\begin{center}
\includegraphics[width=12cm]{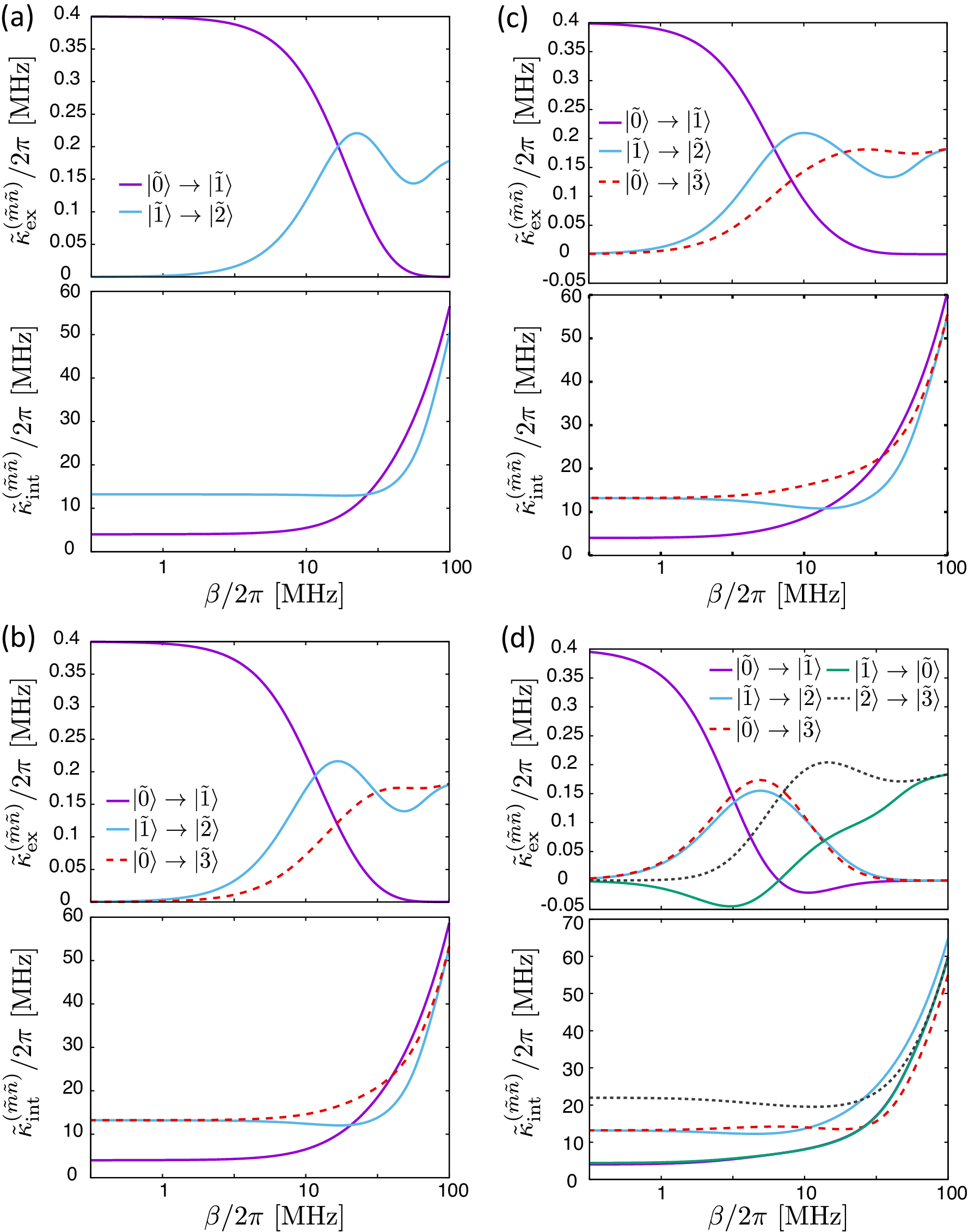}
\end{center}
\caption{
Nominal external and nominal internal decay rates in equation~(\ref{nominal_3_18_21}) for relevant transitions indicated in the panels. Panels (a)--(d) are for $\Delta/2\pi=-7$~MHz (a), 0~MHz (b), 7~MHz (c) and 20~MHz (d), respectively.
The curves for $|\tilde{0}\rangle\rightarrow |\tilde{1}\rangle$ and $|\tilde{1}\rangle\rightarrow |\tilde{0}\rangle$ are almost overlapping in the lower panel of (d).
The other used parameters are the same as figure~\ref{DeltaE_woline_3_16_21}.}
\label{kappa_3_17_21}
\end{figure}

Figure~\ref{kappa_3_17_21} shows the nominal external and the nominal internal decay rates in equation~(\ref{nominal_3_18_21}).
The nominal external decay rate, $\tilde\kappa^{(\tilde{0}\tilde{1})}_{\rm ex}$, decreases to zero as $\beta$ increases.
Some of other nominal external decay rates become finite for $\beta\ne 0$ although they are approximately zero  for $\beta \simeq 0$.
On the other hand, nominal internal decay rate increases rapidly with respect to $\beta$.
Thus, the dips and peaks tend to broaden as $\beta$ increases.
Even if the original internal decay rate of the parametron without a pump field is negligible, 
the nominal internal decay rate increases with the pump strength and eventually exceeds the nominal external decay rate (see \ref{Results for zero internal decay}).

\begin{figure}[h!]
\begin{center}
\includegraphics[width=12cm]{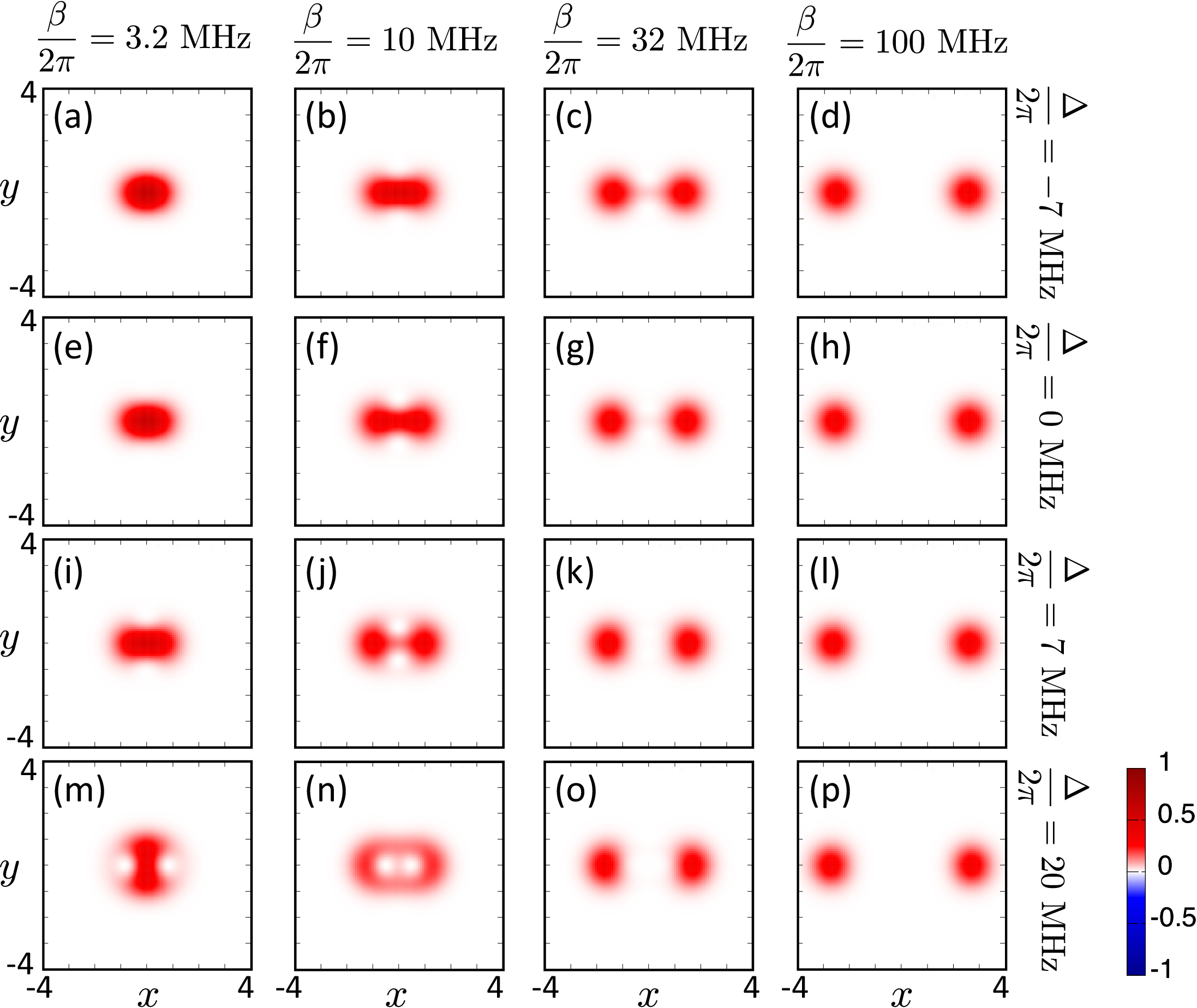}
\end{center}
\caption{
The Wigner function for the stationary state for the values of $\beta$ indicated by the triangles in figure~\ref{eng_3_16_21}(a) and also explicitly written above the panels.
Panels (a--d), (e--h), (i--l) and (m--p) are for $\Delta/2\pi=-7$~MHz, 0~MHz, 7~MHz and 20~MHz, respectively.
The other used parameters are the same as figure~\ref{DeltaE_woline_3_16_21}.}
\label{Wig_3_17_21}
\end{figure}

The  Wigner function in figure~\ref{Wig_3_17_21} illustrates the stationary state for the values of $\beta$ indicated by 
the triangles in figure~\ref{eng_3_16_21}(a).
The profile of the Wigner function depends on the detuning for relatively small value of $\beta$ although it is insensitive to the detuning for $\beta\gg |\Delta|,\chi$ as seen in figures~\ref{Wig_3_17_21}(d), (h), (l), (p).
This is because the stationary state becomes the maximally mixed state of the coherent states, $|\alpha\rangle$ and $|-\alpha\rangle$, with $\alpha \simeq \sqrt{2\beta/\chi}$ for $\beta\gg |\Delta|,\chi$.

\section{Summary}
\label{Summary and discussion}
We have theoretically studied the reflection spectroscopy of a pumped superconducting parametron.
We have developed a method to obtain the reflection coefficient of a parametron and have derived formulae of the reflection coefficient, the nominal external and internal decay rates.
This method can also take into account the effect of the input field beyond the limit of weak input field.
It has been shown that the peak or dip can appear in the amplitude of the reflection coefficient when there is finite difference between the populations of energy levels resonantly coupled by an input field, and the sign of the difference determines whether we have a dip or peak.
We have shown that the nominal internal decay rate increases with the pump strength and eventually exceeds the nominal external decay rate even if the original internal decay rate of the parametron without a pump field is negligible.
The obtained spectrum provides information of the superconducting parametron, such as energy level structure and  amplitude of coupling coefficients between energy levels, and also useful information for calibration of the pump field.

\section*{Acknowledgements}
We thank Y. Suzuki for useful comments. 
This paper is partly based on results obtained from a project, JPNP16007, commissioned by the New Energy and Industrial Technology Development Organization (NEDO), Japan. 
S.M. acknowledges the support from JSPS KAKENHI (grant number 18K03486). 
Y. M. was supported by Leading Initiative for Excellent Young Researchers MEXT Japan and JST presto (Grant No. JPMJPR1919).

\appendix

\section{Hamiltonian of a parametron}
\label{Hamiltonian of a parametron}
\begin{figure}
\begin{center}
\includegraphics[width=5.5cm]{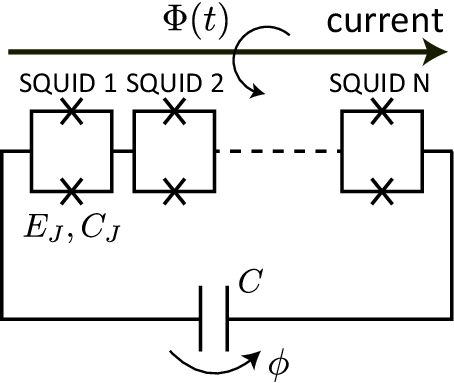}
\end{center}
\caption{
Circuit model of a superconducting quantum parametron consisting of $N$~SQUIDs and shunt capacitor $C$. 
$\phi$ is the overall phase across the junction array.  $\Phi(t)$ is the external magnetic flux threading the SQUIDs.
$E_J$ and $C_J$ are the Josephson energy of a single SQUID and the capacitance of a single Josephson junction, respectively.  
}
\label{KPO_system_6_1_20}
\end{figure}

We derive an effective Hamiltonian for a parametron to make this paper self-contained although it can be found in Ref.~\cite{Wang2019}.
We consider a parametron consisting of a SQUID-array resonator with $N$ SQUIDs depicted in  figure~\ref{KPO_system_6_1_20}.
The effective Hamiltonian of the system is given by
\begin{eqnarray}
H= 4E_C n^2 - NE_J[\Phi(t)] \cos\frac{\phi}{N},
\label{H_KPO_4_16_20}
\end{eqnarray}
where $\phi$ and $n$ are the overall phase across the junction array and its conjugate variable, respectively.
$E_J$ is the Josephson energy of a SQUID. We assume that all the Josephson junctions are identical.
The effective Hamiltonian (\ref{H_KPO_4_16_20}) with a single degree of freedom, $\phi$, is valid provided that $E_J$ is much larger than the charging energy of a single junction~\cite{Frattini2017,Noguchi2020}.
$E_C$ is the charging energy of the resonator, including the contributions of the junction capacitances $C_J$ and the shunt capacitance $C$, and can be experimentally extracted or calculated by finite-element capacitance simulation~\cite{Wang2019}. 
The Josephson energy is modulated as $E_J(t)=E_J+\delta E_J \cos\omega_p t$ by the external magnetic flux, $\Phi(t)$, threading the SQUIDs.

We take into account up to the fourth order of $\phi/N$ in equation~(\ref{H_KPO_4_16_20}) to obtain an approximate Hamiltonian
\begin{eqnarray}
\frac{H}{\hbar} &=& \omega \Big{(} a^\dagger a + \frac{1}{2} \Big{)}
- \frac{\chi}{12} (a + a^\dagger)^4
\nonumber\\
 && + \Big{[} - \frac{N\delta E_J}{\hbar}  +
 2\beta (a + a^\dagger)^2  - \frac{2\chi \beta}{3\omega} (a + a^\dagger)^4 \Big{]}
\cos\omega_p t,
\label{H_9_1_20}
\end{eqnarray}
where $\omega = \frac{1}{\hbar}\sqrt{8E_CE_J/N}$, $\chi=E_C/\hbar N^2$ and $\beta = \omega \delta E_J/ 8E_J$. 
Here, $\beta$ is called amplitude of the pump field in the main text.
$n$ and $\phi$ are related to the annihilation operator $a$ as
$n = -in_0(a-a^\dagger)$ and $\phi = \phi_0 (a+a^\dagger)$
with $n_0^2=\sqrt{E_J/32 N E_C}$ and $\phi_0^2 = \sqrt{2NE_C/E_J}$.
Above, we considered the parameter regime, where $\phi_0/N = 2\sqrt{\chi/\omega}$ is sufficiently smaller than unity so that the expansion of $\cos(\phi/N)$ is valid,
and took into account up to the fourth order of $\phi/N$ to see the effect of the Kerr nonlinearity.
In equation~(\ref{H_9_1_20}), we neglect the last term assuming that $\chi\beta\ll  \omega$, and drop c-valued terms to obtain the following Hamiltonian
\begin{eqnarray}
\frac{H}{\hbar} = \omega a^\dagger a 
- \frac{\chi}{12} (a + a^\dagger)^4
+ 2\beta (a + a^\dagger)^2
\cos\omega_p t.
\end{eqnarray}

\section{Effect of input field}
\label{Effect of input field}
In the main text, we considered the weak input field limit.
The diagonal elements of the density matrix, $\rho^{\rm (F)}_{\tilde{m}\tilde{m}}[0]$, were calculated assuming that they are not changed by the input field, $\Omega=\sqrt{v_b\kappa_{\rm ex}}E$. 
However, we can take into account the effect of finite $\Omega$ by solving the Fourier transform of equation~(\ref{drho_3_18_21}) to obtain the elements of the density matrix.
In the numerical simulations of this section, we assume  $\rho^{\rm (F)}_{\tilde{m}\tilde{m}}[k  \tilde\omega_{\rm in}]= 0$ for $k\ne 0$ and $\rho^{\rm (F)}_{\tilde{m}(\tilde{n}\ne \tilde{m})}[k \tilde\omega_{\rm in}]= 0$ for $k\ne \pm 1$, and take into account from $m=0$ to $5$.

\begin{figure}[h!]
\begin{center}
\includegraphics[width=14cm]{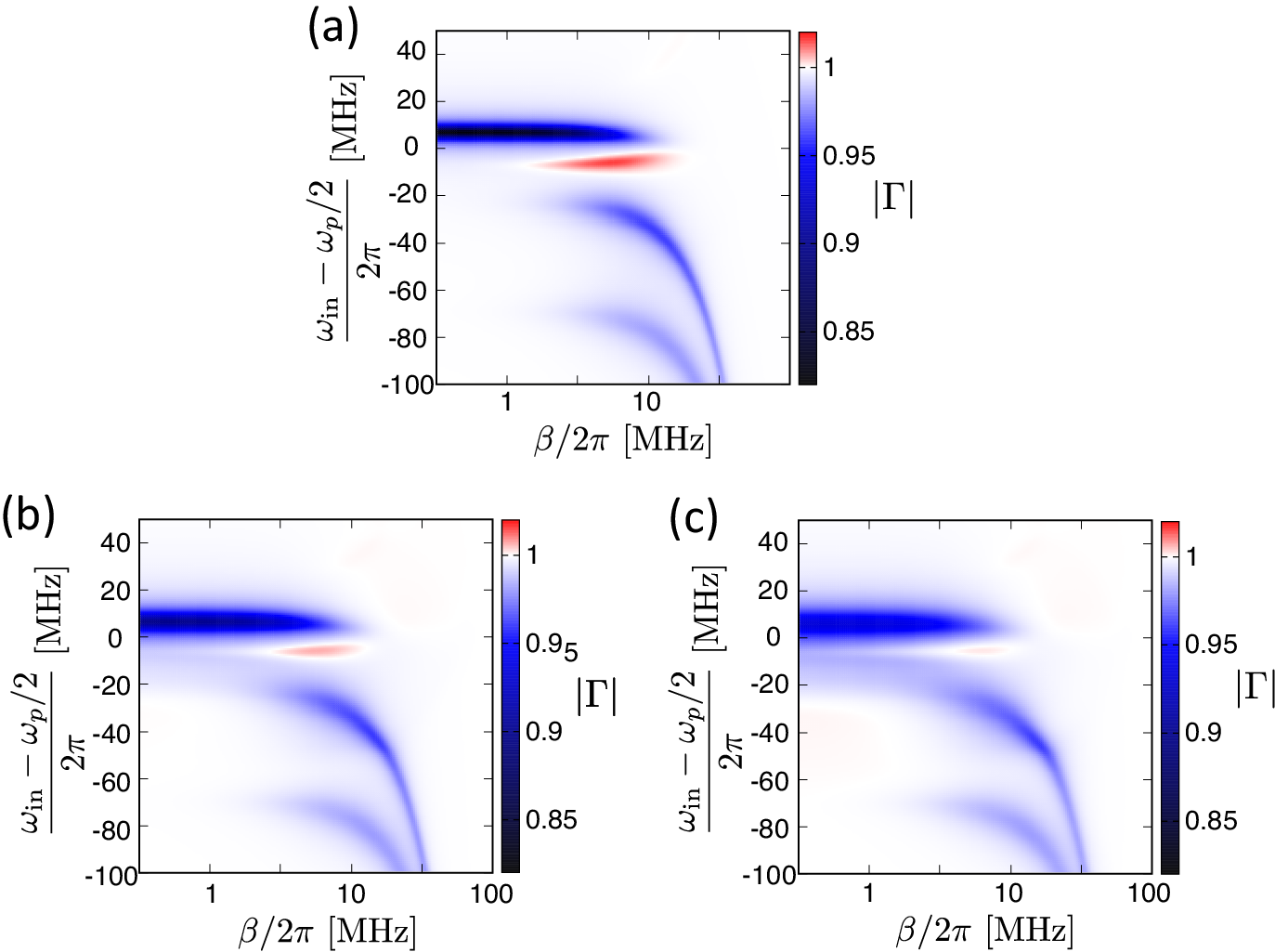}
\end{center}
\caption{
Amplitude of the reflection coefficient as a function of $\omega_{\rm in}$ and $\beta$ for $\Delta/2\pi=7$~MHz with  
$\Omega/2\pi=$1~MHz (a), 2~MHz (b) and 3~MHz(c).
The used parameter set is: $\chi/2\pi = 30$~MHz, $\kappa_{\rm ex}/2\pi=0.4$~MHz and $\kappa_{\rm int}/2\pi=4$~MHz.}
\label{amp_rev_close3_3_29_21}
\end{figure}

Figure~\ref{amp_rev_close3_3_29_21} shows the amplitude of the reflection coefficient in equation~(\ref{Gamma_2_19_21_2}) as a function of $\omega_{\rm in}$ and $\beta$.
The result for $\Omega/2\pi=$1~MHz is approximately the same as the results in figure~\ref{DeltaE_woline_3_16_21} for the weak input field limit.
The peaks (dips) become low (shallow) for larger $\Omega$. 
This is attributed to that $\tilde{\kappa}_{\rm int}^{(\tilde{m}\tilde{n})}-\tilde{\kappa}^{(\tilde{m}\tilde{n})}_{\rm ex}$ defined in equation~(\ref{nominal_3_18_21}) increases as $\Omega$ increases because $|\rho^{\rm (F)}_{\tilde{m}\tilde{m}}[0]-\rho^{\rm (F)}_{\tilde{n}\tilde{n}}[0]|$ becomes small when $\Omega$ is large.

\section{Direct numerical simulations with integration of master equation}
\label{Comparison with results with master equation}
In section~\ref{Weak input field limit}, an approximate formula for reflection coefficient was derived.
The reflection coefficient can be calculated also by a straightforward but time consuming manner.
We integrate the master equation~(\ref{drho_3_29_18}) to obtain the density matrix and calculate $\langle A\rangle [-\omega_{\rm in} + \omega_p/2]$. Equation~(\ref{Gamma_5_25_21}) is  used to obtain the reflection coefficient.
We compare the results with those obtained by the method in section \ref{Weak input field limit}.

In the master equation, we set the initial state of the parametron to the stationary state. 
We integrate the master equation for $0\le t \le 440$ ns with the constant input field.
We used a fourth-order Runge--Kutta integrator with the time step of less than 0.012 ps.

\begin{figure}[h!]
\begin{center}
\includegraphics[width=12cm]{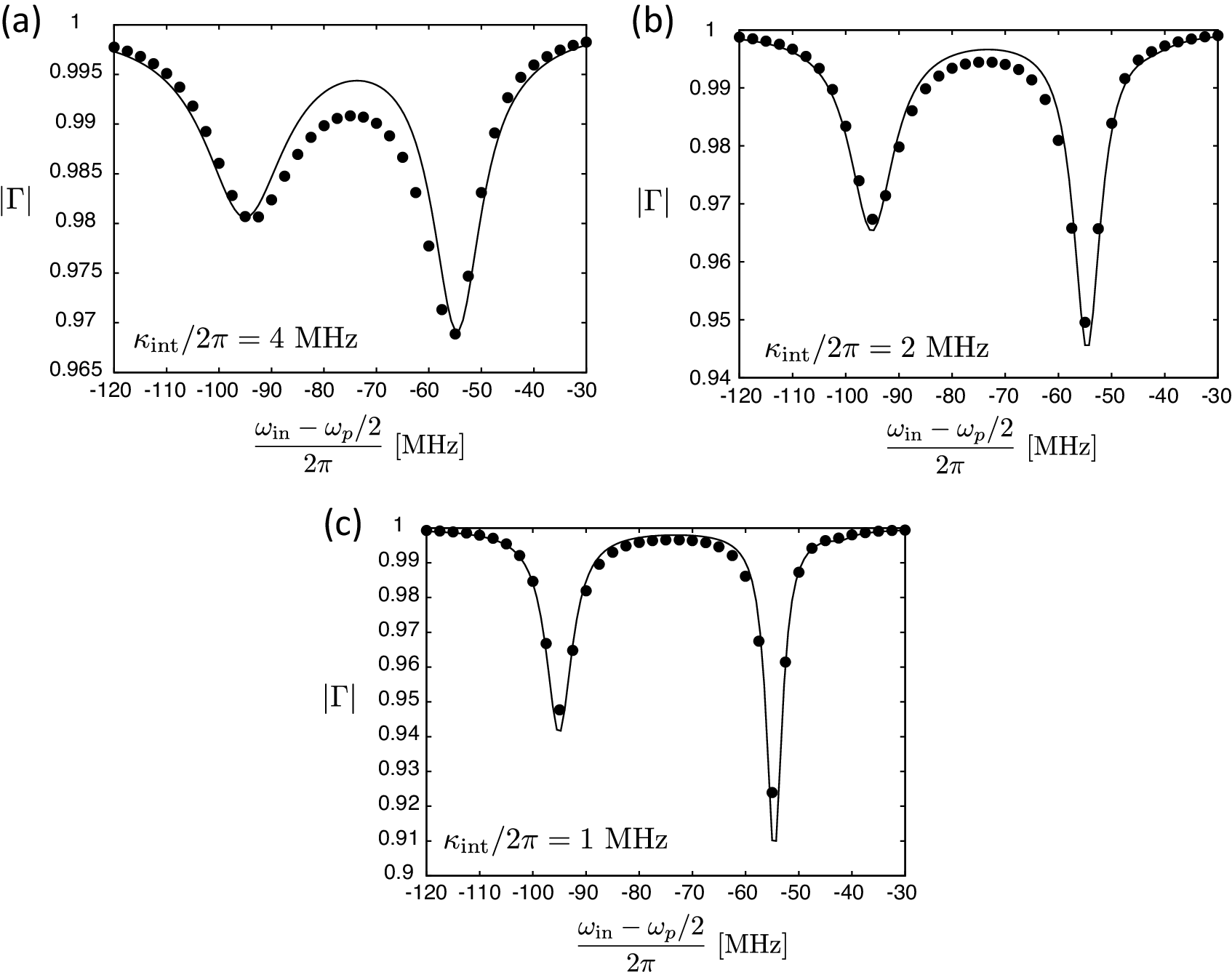}
\end{center}
\caption{
Amplitude of the reflection coefficient as a function of $\omega_{\rm in}$ for $\kappa_{\rm int}/2\pi=4$~MHz (a),
2~MHz (b) and 1~MHz (c).
The used parameter set is: $\beta/2\pi=20$~MHz, $\Delta/2\pi=7$~MHz, $\chi/2\pi = 30$~MHz, $\kappa_{\rm ex}/2\pi=0.4$~MHz and $\Omega/2\pi=8$~kHz.
The solid curves are for equation (\ref{Gamma_3_30_21}), and the circles are for the results obtained by integrating the master equation.
}
\label{amp_com3_8_9_21}
\end{figure}

Figure~\ref{amp_com3_8_9_21} shows the amplitude of the reflection coefficient for the both methods. 
The dip at $(\omega_{\rm in}-\omega_p)/2\pi\simeq -53$~MHz (-94~MHz) corresponds to the transition 
$|\tilde{1}\rangle \rightarrow |\tilde{2}\rangle$ ($|\tilde{0}\rangle \rightarrow |\tilde{3}\rangle$).
It is seen that the results for the method in section \ref{Weak input field limit} approximates well especially near the resonance (dips). 
There is a discrepancy between the two results between the two dips (see figure~\ref{amp_com3_8_9_21}(a)), which we attribute to the fact that we neglect the interference between different transitions in the method in section \ref{Weak input field limit}. 
This discrepancy becomes small when we decreases $\kappa_{\rm int}$ as the dips are well separated as seen in figure~\ref{amp_com3_8_9_21}(b,c).

\section{Results for $\kappa_{\rm int}=0$}
\label{Results for zero internal decay}
The nominal internal decay rate increases with respect to the pump amplitude even if the original internal decay rate of the parametron without a pump field is negligible as shown in this section.
To observe this fact, we consider a fictitious case where $\kappa_{\rm int}$ is zero.

\begin{figure}[h!]
\begin{center}
\includegraphics[width=14cm]{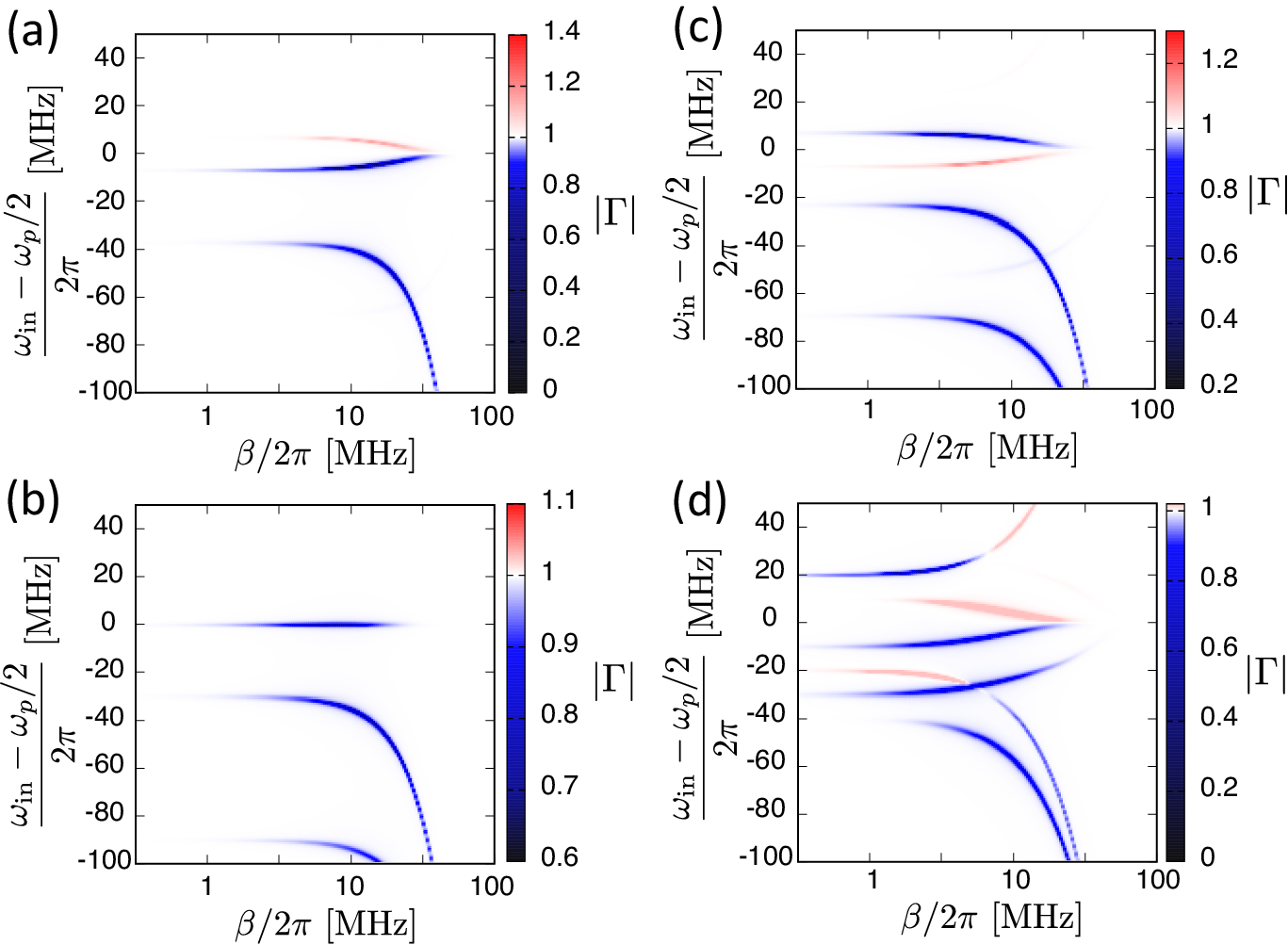}
\end{center}
\caption{
Amplitude of the reflection coefficient in the weak input field limit as a function of $\omega_{\rm in}$ and $\beta$ for $\Delta/2\pi=-7$~MHz (a), 
0~MHz (b), 7~MHz (c) and 20~MHz (d).
The used parameter set is: $\chi/2\pi = 30$~MHz, $\kappa_{\rm ex}/2\pi=0.4$~MHz and $\kappa_{\rm int}/2\pi=0$~MHz.}
\label{amp_rev_close_3_23_21}
\end{figure}

Figure~\ref{amp_rev_close_3_23_21} shows the amplitude of the reflection coefficient calculated for the weak input field limit as a function of $\omega_{\rm in}$ and $\beta$ for $\kappa_{\rm int}=0$.
The dips and peaks are sharper than those in figure~\ref{DeltaE_woline_3_16_21} for $\kappa_{\rm int}/2\pi=4$~MHz.

\begin{figure}[h!]
\begin{center}
\includegraphics[width=12cm]{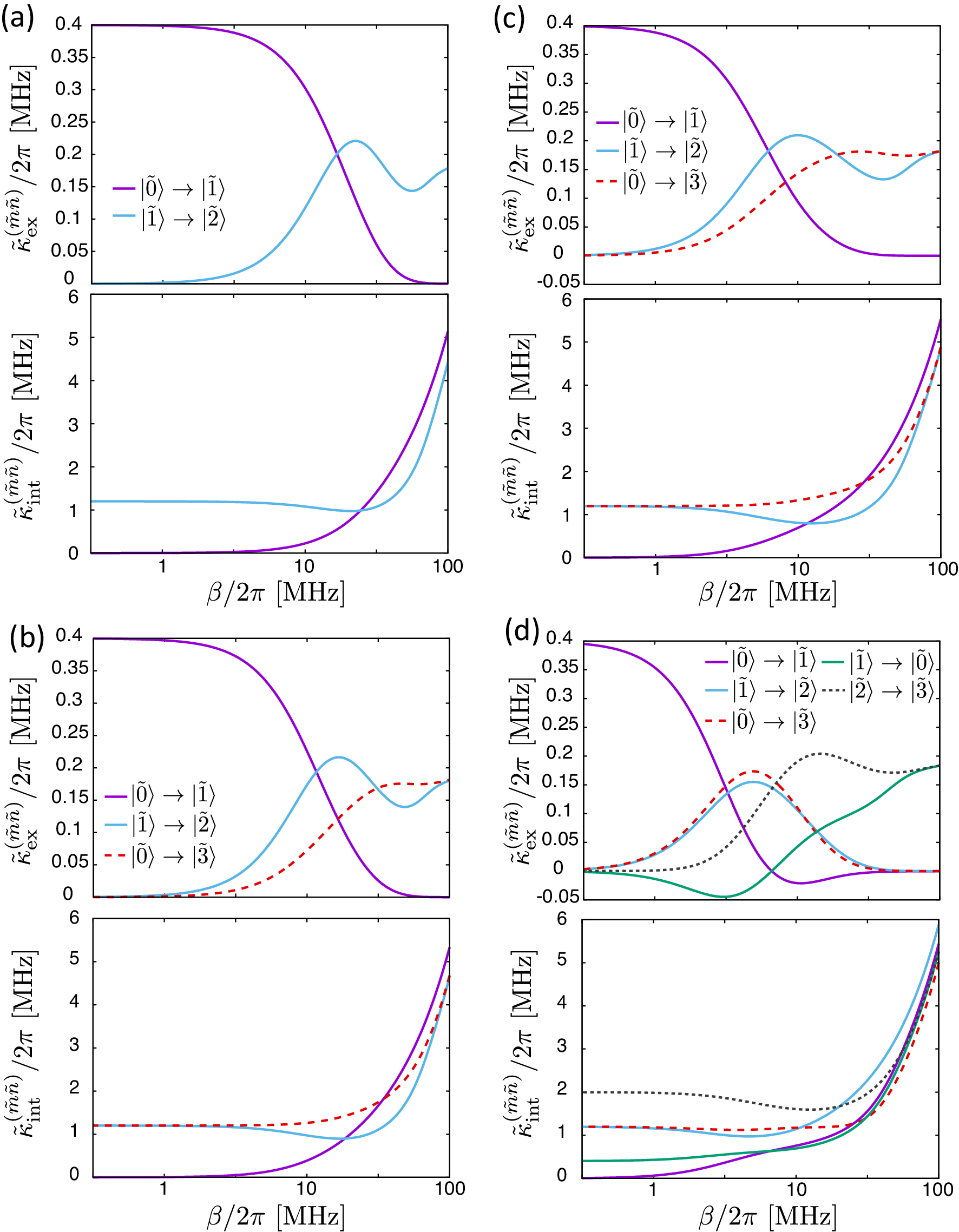}
\end{center}
\caption{
Nominal external and nominal internal decay rates in equation~(\ref{nominal_3_18_21}) for $\kappa_{\rm int}/2\pi=0$~MHz. Panels (a)--(d) are for $\Delta/2\pi=-7$~MHz (a), 0~MHz (b), 7~MHz (c) and 20~MHz (d), respectively.
The other used parameters are the same as figure~\ref{DeltaE_woline_3_16_21}}
\label{kappa_woint_3_22_21}
\end{figure}

Figure~\ref{kappa_woint_3_22_21} shows the nominal external and the nominal internal decay rates in equation~(\ref{nominal_3_18_21}).
The nominal external decay rate is approximately the same as that in figure~\ref{kappa_3_17_21} for $\kappa_{\rm int}/2\pi=4$~MHz.
We attribute this to the fact that the diagonal elements of the density matrix are approximately the same in both cases.
On the other hand, nominal internal decay rate is much smaller than that in figure~\ref{kappa_3_17_21} due to vanishing $\kappa_{\rm int}$.
However, the nominal internal decay rate increases rapidly with respect to $\beta$ because  $Y_{\tilde{m}\tilde{m}}$ and   $Y_{\tilde{n}\tilde{n}}$ increase with $\beta$.
Thus, the broadening of the dips and peaks occur also in the case of zero $\kappa_{\rm int}$.

\end{document}